 \documentstyle[12pt,aasms4]{article}

\def\etal{{\it et~al.\ }}
\def\eg{{\it e.g.\ }}
\def\ie{{\it i.e.\ }}
\def\kms{km s$^{-1}$}
\def\ergs{erg s$^{-1}$}
\def\ana{Astron. Astrop.}

\def\minmag{\lower.5ex\hbox{$\; \buildrel < \over > \;$}}
\def\magmin{\lower.5ex\hbox{$\; \buildrel > \over < \;$}}
\def\gtwid{\mathrel{\raise.3ex\hbox{$>$\kern-.75em\lower1ex\hbox{$\sim$}}}}
\def\ltwid{\mathrel{\raise.3ex\hbox{$<$\kern-.75em\lower1ex\hbox{$\sim$}}}}
\def\ref{\par\noindent\hangindent.5in\hangafter=1}
\newcommand{\beq}{\begin{equation}}
\newcommand{\eneq}{\end{equation}}
\newcommand{\beqar}{\begin{eqnarray}}
\newcommand{\eneqar}{\end{eqnarray}}
\newcommand{\barn}{\begin{eqnarray*}}
\newcommand{\earn}{\end{eqnarray*}}
\lefthead {Done and Krolik}
\righthead {BLR Kinematics in NGC 5548}

\begin{document}
 \footnotesize
\title{Kinematics of the Broad Emission Line Region in NGC 5548}

\author{Christine Done}

\affil{Department of Physics, Durham University, South Road, Durham DH1
3LE, UK}

\and

\author{Julian H. Krolik}

\affil{Department of Physics and Astronomy, Johns Hopkins University,
	   Baltimore, MD 21218}

\begin{abstract}

    We derive both total flux and velocity-resolved response functions for
the CIV 1549 emission line from the data obtained in the 1993 NGC 5548
monitoring campaign.  These response functions imply: 1.) the emission
region stretches from inside 1 lt-d to outside 10 lt-d, and is
probably better described as round than flat;
2.) the velocity field is dominated by
a red/blue symmetric component (e.g. 2--d or 3--d random motions, or
rotation in a disk) but there is also {\it significant} radial infall.
Quantitative modelling indicates that the random speeds are typically a few
times as large as the radial speed. However,
no simple model gives a completely acceptable fit to the data.
These inferences rule out numerous simple and otherwise plausible models
for broad line region dynamics, including outflowing
winds, radial free-fall, rotation in a disk, or collisionless orbital motion.

\end{abstract}

\section{Introduction}

    One of the most salient characteristics of active galactic nuclei is their
strong, broad emission lines.  Ever since the discovery of these objects
(\cite{S43}), the origin of the large velocities implied by these widths
has been the subject of much discussion and speculation (\eg \cite{MC85}).
In addition to the
intrinsic interest in answering this question, there has also long been hope
that if we could learn the nature of the motions in the broad emission line
region, we could use that material as diagnostics of the accretion flow
generally believed to power the central engine.

Despite the many reasons to investigate the kinematics of the broad emission
line gas, there remains considerable uncertainty about their true nature.
Early work that tried to determine the kinematics of the BLR relied on
fitting profile shapes to particular velocity models
(for example, infall, outflow, random,
disk etc.). However, this technique has difficulty producing unambiguous
results
because there is too much freedom to change the distribution of physical
conditions, and hence the distribution of line emissivity (\cite{S78};
\cite{C80}; \cite{W81}; \cite{M82}; \cite{W84}; \cite{Ka93}).

Velocity-resolved reverberation mapping is potentially a much more powerful
method (\cite{BM82}; \cite{WH91}; \cite{P92}).  This method rests
on a comparatively clean theoretical base: If the emission lines are
powered by photoionization due to the continuum radiated by the central engine,
the emission line light curves should be delayed and smoothed replicas of
the continuum light curve.  This is because the light travel time from
the continuum source to the emission line region [$\sim (1$ -- $100)
L_{44}^{1/2}$ lt-d, where $L_{44}$ is the ionizing continuum
luminosity in units of $10^{44}$ \ergs] is likely to be significant both on
the human timescale and with respect to the timescale of continuum
variations, while the local response times within the line-emitting gas are
probably many orders of magnitude shorter.  In principle, then, one should
be able to unfold the fluctuation histories of the continuum and the
emission lines in order to map the line response onto the paraboloidal
surfaces of constant delay. By doing this separately for segments of
the lines with different line of sight velocities, one obtains a picture
of the correlation between velocity and position.  Qualitatively,
whichever side of the line responds with the least delay
to continuum fluctuations is the one predominantly made on the near
side of the source; \ie if the red wing leads, infall prevails, while
if the blue wing leads, the flow is dominated by outward motion.  If both
sides move together, there is little net radial motion.

However, to fully recover a map of the velocity-resolved line
response requires large amounts of high signal/noise data well-sampled
on the relevant timescales.  Such campaigns require immense observing effort.
For this reason, the only attempt at this program made by anyone hitherto
was the work by \cite{WH94}, who used ground-based monitoring
data on the type 1 Seyfert galaxy NGC 3516 to find the velocity-resolved
response of the H$\alpha$ line.  While they were able to demonstrate that
the motions could not be predominantly radial, they were hampered by
limited sampling.  Their data included only 18 measurements of the line
profile, and these were spaced irregularly with a mean interval of $\simeq 9$d,
while the response was almost entirely more rapid than $\simeq 20$d.

Because poor temporal sampling of the lightcurves is very common, many
have pragmatically adopted a more limited goal:
to recover only a weighted (and biassed: \cite{EK88}) moment of the
response function (the characteristic
lag) by studying the cross-correlation of the continuum fluctuations with the
line fluctuations, and dividing the line profile into only two segments,
the red and blue halves.
Those studies with the least sparse data also support the idea that
radial motions are not the dominant velocity field, most notably in
NGC4151 (\cite{C87}; \cite{G88}; \cite{M91}), NGC5548 (\cite{KG91a};
\cite{KG91b}; \cite{K95}) and Mkn 279 (\cite{S94}).  However, many of
these studies (particularly the earlier ones) also
suffered from inadequate temporal resolution.  In
addition, there have been conflicting claims regarding the same objects, most
notably the subject of this paper, NGC 5548. \cite{RM90} favored
radial outflow with obscuration; \cite{CB90} argued for radial infall;
\cite{KG91a} saw evidence for primarily random motions; while \cite{M93}
interpreted the data as showing bi-conical infall.

The 1993 monitoring campaign on NGC 5548 (\cite{K95}) provides
the best dataset so far from which to recover the full velocity-resolved
response function.  That campaign obtained 39 UV spectra of the nearby ($z
= 0.0174$) type 1 Seyfert galaxy NGC 5548 with the {\it HST}, taken at
1d intervals.  In coordination with the {\it HST} observations, {\it
IUE} spectra were obtained every 2d, beginning 36d before the first
{\it HST} spectrum, and continuing to the end of the {\it HST}
campaign.  Because (as we shall show) there is significant response at
lags up to at least $\simeq 20d$, the continuum fluxes recorded with
{\it IUE} give substantial aid to the analysis of the {\it HST}
emission line light curves.

Two main considerations limit the choice of emission lines for
velocity-resolved reverberation mapping.  First, stronger lines are
preferred so that the S/N remains high even when they are divided into
segments.  Because stronger lines carry a larger portion of the total
emission line flux, they are also more ``representative" of
the emission line region as a whole.  Second, it is desirable that
any absorption features or
different lines blended into the main feature be as weak as possible.  Applying
these criteria to the NGC 5548 spectrum identifies CIV 1549 as by far
the best line to use.  It is the strongest single line in the spectrum,
exceeding even Ly$\alpha$ in mean flux. In addition, the only significant
blending in CIV 1549 is a small contribution to its extreme red wing by
HeII 1640, although it does have two weak absorption features
a few hundred km/s to the blue of the systemic redshift (\cite{K95}).
By contrast, Ly$\alpha$ is strongly contaminated in its blue
wing by geocoronal Ly$\alpha$, in its red wing by NV 1240, and is
strongly absorbed.  The next
stongest line after Ly$\alpha$ is CIII] 1909, but its flux is only $\sim 0.2
\times$ that of CIV 1549 (\cite{K95}), and it is also blended with
SIII] and AlIII.  For these reasons, we
concentrate exclusively on CIV 1549 in the present work.

\section{The Data}

  In the following analysis we use two data sets: the {\it HST} data by itself,
and the {\it HST} data combined with the {\it IUE} data.  We use the combined
data set only for the continuum lightcurve; all CIV 1549 data that we use is
taken solely from {\it HST} observations.

The statistical errors associated with the {\it HST} measurements are extremely
small, around $\sim 0.3$\% {\it rms} for the total flux in CIV, and
$\simeq 1.7\%$ for the continuum (\cite{K95}).  However, the known {\it
systematic} errors in the
repeatability of FOS short timescale (orbit to orbit) photometry are
at least 1.4\% {\it rms}, from analysis of well centered, standard star
calibrations (\cite{K95}).
Not all the spectra of NGC5548 are well centered, so this is
a lower limit to the size of the systematic errors. The true
size of the systematic error in these data is not well known:
\cite{K95} estimate that it lies within the
range of $2-4.5$\%.
As these are systematic errors they are not independent
(the continuum and all the CIV velocity-resolved line components
should be affected in the same way in a given spectrum, but are uncorrelated
with respect to systematic errors in other spectra).  Because there is no
obvious way to take account of this, we add an error of 2\% in quadrature
to the continuum, total CIV, and velocity-resolved
lightcurves.  Note that this procedure leads to a slight overestimate of
the amount of correlation between the lightcurves at zero lag, but this
overestimate should be independent of velocity.
Errors for the {\it IUE} data are rather larger: typically $\simeq 6\%$,
but with a large dispersion from spectrum to spectrum.

   When we discuss the total CIV line flux, we use the figures given in
\cite{K95}.  For velocity-resolved work, we divided the CIV profile
into four segments of approximately equal mean flux; we call them the
blue wing, blue core, red core, and red wing.  ``Core" denotes the range
from zero velocity relative to systemic out to 2456 \kms; ``wing" denotes
an integration from 2456 \kms to 10840 \kms.  In each case, the flux was
obtained by a direct integration from the measured spectra; no special
correction was made for the narrow absorption feature which appears
within the blue core. This particular velocity cut has the advantage that
it gives roughly equal flux in all the line segments.
These data and integrations were kindly made available
to us by Kirk Korista.

The four lightcurves are generally remarkably similar to each other,
showing immediately that the velocity field must be predominantly
red/blue symmetric (\cite{K95}).  Table 1 gives quantitative
measures for testing just how similar the lightcurves are.  For the
total line flux, total continuum flux, and each of the four line segments,
it shows: the mean, $<F>=
(1/N)\Sigma_j F_j/\sigma^2_j$; total variance (including the effect of
the error bars) $\sigma_{tot}^2=1/(N-1)\Sigma_j (F_j-<F>)^2$; error
variance, $\sigma_{err}^2=N/\Sigma_j\sigma^{-2}_j$;
the fractional intrinsic {\it r.m.s.} variability relative to the mean,
$\delta F/<F>=\sqrt{(\sigma_{tot}^2-\sigma_{err}^2)/<F>^2}$;
and the signal/noise ratio $S/N =\delta F/\sigma_{err}$.

The dilution of the core components by the (constant) narrow line
flux can be estimated  in two ways. Firstly, from
a previous {\it HST} observation of NGC5548
in which the source was in an unusally low state so that the
narrow lines were especially prominent (July 1992, \cite{cbw93}).
This gave an estimated  narrow line flux of $7.2\pm
1.2\times 10^{-13}$ ergs cm$^{-2}$ s$^{-1}$, compared to the
mean red plus blue core lightcurve flux in these data
of $3.8\times 10^{-12}$ ergs cm$^{-2}$ s$^{-1}$, i.e. $19\pm 3$\%
of the total red plus blue core flux. Secondly,
fitting the mean {\it HST} spectrum gives an
(absorption corrected) estimate for the narrow line contribution of
$\sim 5.6\times 10^{-13}$ ergs cm$^{-2}$ s$^{-1}$ on a total core flux
of $3.75\times 10^{-12}$ ergs cm$^{-2}$ s$^{-1}$, i.e. a 15\%
contaminating flux. Systematic errors in modelling the line and
absorption features dominate over statistical errors, but clearly
we expect the dilution to be {\it at least} 10\%.
Thus for the two core components, the fractional {\it r.m.s.} variability
has been adjusted to allow for a 10\%
contribution to the mean flux due to the narrow component.

There is a systematic trend for both $\delta F/<F>$
and $S/N$ to rise monotonically
from the blue wing across the profile to the red wing.  This trend is strong
enough that the relative source
variance of the red wing is significantly different from that
of the blue wing (95\% confidence on an F test -- ratio of variances --
with 39 degrees of freedom).  Some of the additional variance in the red
wing may
be due to contamination from the blue wing of HeII (\cite{K95}),
but, as we argue in \S 5.1, we do not believe that this contamination
can be large enough to explain much of the difference.
Even allowing for only 10\% dilution for the narrow line flux on
the red and blue cores, they too have significantly ($\ge 90$ \% confidence)
more variance than the blue wing.

Thus the line lightcurves on their own show that there are subtle but
significant variability differences between the velocity components, in
the sense that there is more variability in
the red than the blue side of the line.  We will elaborate on how
to interpret that contrast in \S 5.3.

Finally, we note that because the variance is predominantly due to
fluctuations on the longest timescales, the fluctuations seen
on timescales of 1 -- 2d are largely noise.  For this reason, when we
construct lightcurves with 2d intervals from the {\it HST} data, we do so
by convolving the 1d sampling with a Gaussian kernel $\propto \exp[-(\Delta t/
1$d$)^2]$.

\section{Response Functions}

\subsection{Methodology}

     Our fundamental assumption is that the line flux fluctuations
at time $t$ are due to earlier fluctuations in the continuum flux.  We write
the relation between the flux $F_l (t,u)$ in line $l$
at time $t$ and line of sight velocity $u$, and the
continuum flux $F_c(t)$ in the form
\beq
F_l (t,u) = \langle F_l \rangle + \int_0^{\tau_{max}} \, d\tau \,
\Psi_l (\tau, u)\left[F_c (t - \tau) - \langle F_c \rangle (\tau)\right] ,
\eneq
where $\tau$ is the time delay, the angle brackets denote a time average,
and $\Psi_l(\tau,u)$ is called the ``response
function" for velocity $u$ of line $l$.
As explained in \cite{KD95}, the proper time average of the continuum
flux is a function of $\tau$. As is also explained in \cite{KD95},
we use the convolution equation between the line and continuuum
{\it fluctuations} rather than between the total line and continuum flux
as the line response need not be truly linear (see also \cite{H94}).

Equation 1 is an appropriate approximation when the fractional continuum
fluctuations are small, i.e.
\beq
{\delta F_c \over \langle F_c \rangle} < 1.
\eneq
In addition, the error in the linearized description of the line response is
dominated by uncertainties in the data rather than nonlinear corrections
to the model when
\beq
\max{ \left[{\sigma_l \over \langle F_l \rangle},{\sigma_c \over \langle F_c
\rangle } {\partial \ln F_l \over \partial \ln F_c}\right]} > {1 \over 2}
{\partial^2 \ln F_l \over \partial (\ln F_c)^2}\left({\delta F_c \over \langle
F_c \rangle}\right)^2 .
\eneq
Here $\delta F_{c,l}$ and $\sigma_{c,l}$ are the rms fluctuations
and rms error in the continuum and line, respectively.
Photoionization models suggest that for CIV 1549 the first logarithmic partial
derivative is $\simeq 1$ in the range of ionization parameters
likely to apply in NGC 5548.  For example, \cite{K91} found that
in NGC 5548 $\Xi$ was in the range 0.1 -- 0.6, corresponding (for
the specific continuum shape they chose) to $U \simeq
0.003$ -- 0.04; their models indicated $\partial \ln F_l/\partial\ln F_c
\simeq 1.1$.  Similarly, using a variety of continuum shapes,
\cite{ogg95}, \cite{bin89}, and \cite{fm82}, predicted that when
$U \sim 0.01$ -- 1, $\partial \ln F_l/\partial\ln F_c \simeq 1$ and
the second logarithmic partial derivative is $\simeq -0.4$ -- $-1$.
Considering the {\it HST} data alone, $\delta F_c /\langle F_c \rangle \simeq
0.14$, $\sigma_c /\langle F_c \rangle
= 0.026$, and $\sigma_l / \langle F_l \rangle \simeq 0.02$ for the total
CIV flux; we have included the systematic error in these numbers.  However,
when we form the merged continuum light curve and smooth
the {\it HST} CIV light curve to 2d resolution, these
quantities become $\delta F_c /\langle F_c \rangle \simeq 0.24$, $\sigma_c
/\langle F_c \rangle
= 0.05$, and $\sigma_l / \langle F_l \rangle \simeq 0.013$.
Thus, the first condition is easily satisfied, and the second condition is
likewise (weakly) satisfied due to the relatively large systematic errors, and
the large random errors of the {\it IUE} measurements. However, without the
systematic errors, the second order corrections would start to dominate.
When that is the case, while the linear approximation may still account
for much of the system's behavior, it is no longer possible
to find purely linear response models which would fit
the lightcurves to within the errors.  We caution that
progressively better data will therefore require more complex techniques
which take into account nonlinear line response (\cite{H94}; \cite{fP94}).

We solve the convolution equation by the method of regularized linear
inversion (\cite{KD95}).  Deconvolution is inherently a
noise amplifying process, so merely minimizing the $\chi^2$ goodness of
fit between the observed line and that predicted from convolving the
continuum with some transfer function leads to a derived transfer function
which is extremely ``choppy". The linear regularization constraint
minimizes the {\it sum} of $\chi^2$ and a measure of how far the inferred
transfer function differs from some {\it a priori} smooth form.  For the
results presented here, our smoothness constraint is minimal deviation
from either a linear or parabolic form.
To control the relative weights given to minimizing
$\chi^2$ and the quantity measuring deviation from smoothness, we multiply
the smoothness measure by a Lagrange multiplier we call $\lambda_*$.
When $\lambda_* = 1$, equal weight is given
to the two quantities to be minimized; $\lambda_*=0$ removes the {\it a
priori} term so that only $\chi^2$ is minimized, while $\lambda_*\gg 1$
gives much more weight to the {\it a priori} expectations of the
transfer function shape than to the discrepancies between the light
curve it predicts and the observed light curve.  To gauge the quality
of a solution, we compare the $\chi^2$ it predicts to the effective
number of degrees of freedom, a quantity which reflects the number of
independent data points adjusted by the relative S/N of the line
and continuum light curves (\cite{KD95}).

To estimate the uncertainty
in the derived transfer function due to measurement error,
we resample both the line and continuum light curves within the
error distribution 100 times, and then calculate the {\it r.m.s.}
dispersion at each lag of the resulting 100 derived transfer functions.
It is important to recognize that this gives the range of
uncertainty {\it constrained by the underlying assumptions of the model},
{\it i.e.\ } the particular values of $\lambda_*$ and the maximum
lag $\tau_{max}$,
and also that the error bars so derived are not independent.
For a comparison of the properties of
this inversion technique with others, such as maximum entropy and subtractive
optimized local averages, see \cite{H94}; \cite{K94}; \cite{fP94}.

\subsection{Interpretation}

Unfortunately, response functions produce a map of line marginal
emissivity not as a function of position relative to the source, but
as a function of delay with respect to our line of sight.  As a result,
the surfaces of projection are oriented in an arbitrary direction (ours),
and cut across a wide range of radii---the surface of delay $\tau$ includes
matter at all radii $ \geq c\tau/2$. Interpretation of the shape of the
derived response function therefore involves a certain amount of
model-dependence.

   To see how this model-dependence enters, we begin by writing the
line luminosity at line of sight velocity $u$ in terms of local properties
inside the emission line region:
\beq
L_l (t,u) = \int \, d\tau \, \int \, d^3r \, n({\bf r}) A({\bf r})
S_l\left[{\bf r},F_{ion}\left(t - \tau\right)\right]
\delta\left[\tau-\tau({\bf r})\right]
\int \, d^3v \, f({\bf v})\delta\left(u-{\bf v \cdot z}\right),
\eneq
where $n({\bf r})$ and $A({\bf r})$ are the number density and surface area
of the BLR clouds, $S_l$ is their surface brightness in the line,
$F_{ion}$ is the local continuum flux, $f({\bf v})$ is the clouds'
velocity distribution, and the delay as a function
of position is
\beq
\tau({\bf r}) = {r \over c}\left(1 - \cos\theta\right)
\eneq
for clouds whose angle to the observer's line of sight is $\theta$.
The two $\delta$ functions pick out the component of velocity in the
direction of the observer and the spatial positions corresponding to a given
lag.  This description also assumes that the clouds radiate line photons
isotropically; it is not hard to modify equation 4 in order to include
anisotropic radiation, but it is an unnecessary complication here.

   Using the linear response approximation, and assuming that the continuum
is radiated isotropically, we can approximate the line
luminosity fluctuations by
\beq
\delta L_l (t,u) = \int \, d\tau \,
\int \, d^3r \, n({\bf r}) A({\bf r}) {\partial S_l \over
\partial F_{ion}}{\delta L_c (t - \tau) \over 4\pi r^2}
\delta\left[\tau-\tau({\bf r})\right]
\int \, d^3v \, f({\bf v})\delta\left(u-{\bf v \cdot z}\right).
\eneq
{}From this form it is clear that the response function is
\beq
\Psi_l (\tau,u)=\int \, d^3r \, n({\bf r}) {A({\bf r})\over 4\pi r^2}
{\partial S_l({\bf r})\over \partial F_{ion}}
\delta\left[\tau-\tau({\bf r})\right]
\int d^3v f({\bf v}) \delta(u-{\bf v \cdot z}).
\eneq
 Integrating $\Psi_l (\tau,u)$ over all velocities $u$ gives the
total line response.

    The most important element of model-dependence is that a geometric
symmetry must be assumed in order to use the response function as defined
by equation 7 to infer a map of the emissivity.  Because it is the
simplest assumption (and also because the lines in NGC 5548 have such
large equivalent widths that their sources must cover a fair fraction of
$4\pi$ around the continuum: \cite{K91}), we assume spherical symmetry.
Equation 7 integrated over line of sight velocity then reduces to:
\beq
\Psi_{tot}(\tau)=\int^{r_{max}}_{c\tau/2} \, dr \, n(r) A(r)
{\partial S_l(r)\over \partial F_{ion}} {c\over 2r}
\eneq
Thus, for a thin shell at distance $r_o$ from the source, the total response is
constant at $c/(2r_o)\ n(r_o) A(r_o) \partial S_l(r_o)/\partial F_{ion}$
for all lags $0\le\tau\le 2r_o/c$, i.e. all segments of the shell contribute
equally to the response function, but at a lag $\tau=(1-\cos\theta)r_o/c$.

Two important consequences follow from the fact that in this model
each radial shell produces
a square wave response function.  First, if this
model applies, the marginal emissivity distribution may be directly read off
the total flux response function.
Second, while different shells stop contributing to the response function at
different maximum lags, they all contribute at $\tau = 0$.  Therefore,
if $\partial S_l/\partial F_{ion} > 0$ everywhere, the response
function always peaks at zero lag.  If there is no material inside $r =
c\Delta\tau/2$, where $\Delta\tau$ is the minimum interval at which the
response function may be found, $\Psi_{tot}(\Delta\tau) = \Psi_{tot}(0)$;
if there is some material inside that point, the zero lag point is an absolute
maximum.

A few further comments apply to the interpretation of velocity-resolved
response functions.  While qualitative information can be gleaned from their
direct comparison, quantitative measures do not follow simply from their
form.  Therefore, after computing the velocity-resolved response functions
for these data, we will also construct parameterized models whose
predictions may be directly compared with the light curves.  We
will then find the parameters for these models which, given the observed
continuum light curve, minimize $\chi^2$ for the predicted velocity-resolved
line light curves.

\section{Total CIV Response}

Considered on its own, there are $N=39$ points spanning 38d in the
{\it HST} data.  The maximum lag length $M$ that can be examined is
then $M=N/2=19$, so we can search for response functions with a maximum
lag $\tau_{max} = 18$d.  However, if we make this choice for $M$, in
the limit of $\lambda_* \ll 1$ there
would be as many free parameters as line measurements,
so a solution could always be found, even if there were in reality
{\it no} connection between the continuum lightcurve and the line lightcurve.
On the other hand, when $\lambda_* \gg 1$, the response function is effectively
being fit by a model with many fewer free parameters---two for a linear
smoothing constraint, three for a quadratic condition.
Thus the number of degrees of freedom in that limit becomes effectively 16
or 17.  Even with extremely large amounts of smoothing ($\lambda_* \sim 10^4$),
the data always give $\chi^2 \ll 16$ (for $\lambda_* = 10^4$ and
quadratic regularization, the
actual reduced $\chi^2 = 0.24$), showing that the
{\it HST} data alone are consistent with a very smooth response function
(Fig. 1).

However, with this length of response function we are throwing
away half of the line data, and the discarded data are the ones
with the most variability.   A more reasonable length to examine is
$M=N/3=13$, or $\tau_{max} = 12$d, so that in effect
the solution is tested on at least as much data as it is derived from,
and some of the more rapid variability is included in the fit.
This choice gives an excellent solution with $\chi^2_\nu=0.6$ for
$\lambda_*=1$, or $\chi^2_\nu=1$ for $\lambda_*=750$ (Fig. 2).
Even shorter response functions can give an acceptable
solution, down to $M=10$, where $\chi^2_\nu=1$ can be found with $\lambda_*=0$.
Thus the {\it HST} data alone {\it require} that there is non-zero
CIV response over lags at least from 1 to 10d.  However, these shortest
possible response functions are not necessarily the best estimates of
the response function.  In particular, as Fig. 2 illustrates, the
shorter the permitted response function, the greater the effect of
the correlated error at zero lag due to the systematic calibration errors.

  Additional information is provided by the {\it IUE} continuum lightcurve.
Because the {\it IUE} data begin 36d before the {\it HST} data, use of
the {\it IUE} continuum light curve enables us to use the entire {\it
HST} line light curve even while searching for response functions that
extend to lags as great as 36d.  The {\it IUE} data also carry another
benefit: the continuum was substantially more variable during the time
only {\it IUE} was monitoring NGC 5548 than during the time of the {\it HST}
observations.  Clearly we cannot now resolve any
structure on scales shorter than 2d, but, as we have already argued, any
signal in the data on such short timescales is probably noise-dominated
in any case.

   Using the combined data set for the continuum light curve,
we find that with $\tau_{max} = 36$d the response can be well fit by a
quadratic function, giving $\chi^2_\nu \ll 1$ even for $\lambda_*\gg 1$
(Fig. 3).  To show the non--uniqueness inherent in the deconvolution
procedure, but also demonstrate which features of the response function
are robust, we present Fig. 4, which displays the derived response
with $\lambda_*=1$  for $\tau_{max}$ between 16d (the smallest for which
$\chi^2_\nu\sim1$ with $\lambda_*=1$) and 36d.  Two features
are independent of detailed choices: the total flux response function
peaks at zero lag, and it drops to zero by $\simeq 20$d.  Since we
fix $\lambda_* = 1$ for this set, the $\chi^2$ falls as $\tau_{max}$
increases, but this is at least partially due simply to the larger
number of parameters available.  For future reference, we call the
solution with $\lambda_* = 10^4$, a quadratic regularization condition,
and $\tau_{max} = 36$d the ``minimal structure" solution (Fig. 3), and the
one with $\lambda_* = 1$, a linear regularization condition, and
the same $\tau_{max}$ the ``maximal structure" solution (Fig. 4, solid line).
The former
can be viewed as the smoothest solution which produces an acceptable
$\chi^2$, while the latter can be viewed as the solution with the
greatest amount of plausible detailed structure.

Several qualitative points follow directly. First, the fact that the response
function peaks at $\tau = 0$d demonstrates that there
is either considerable line-emitting gas on our line of sight, or
that there are large amounts of gas at radii smaller than
1 lt-d, together with more extended material which gives the
non--zero response at $\sim$ 10 lt-d. A possible geometry for this latter
case is a disk of material with an inner radius of less than
1 lt-d (see e.g. the transfer functions of \cite{WH91}; \cite{P92}).
However, in order for a
disk to produce the observed broad line profiles requires that there
is a component of the velocity in the line of sight, i.e. that the
disk is not viewed face on.
This seems rather implausible for Seyfert 1's in general given current
Seyfert unification schemes which postulate that these objects are
seen at rather small inclination angles
(e.g. the review by \cite{A93}), and for NGC 5548 in particular,
which has a strong Compton reflection signature in its X--ray spectrum
(\cite{NP94}; \cite{D94}).  A simple thin disk is also
strongly ruled out by the velocity field (see \S 6.2). Thus we prefer the
former option, in which there is gas along our line of sight.

In order for there to be nothing special about the direction to us, we
infer that the gas is probably distributed quasi-spherically.  This
inference is supported by the rather large equivalent widths of the
lines in this object (\cite{K91}).  Large response from gas on the
line of sight also means that the surface brightness on the outside of
the clumps cannot be much less than the surface brightness on the side
facing the continuum source.  This is consistent with standard
photoionization models, which predict that the front/back ratio for
CIV 1549 even when the clouds are very optically thick in Ly$\alpha$
is only $\simeq 3$ (e.g. \cite{KK86}; \cite{F92}).

Second, there is significant response over all lags from 1d to $> 16$d.
Indeed, we have found that when the maximum lag considered falls below 12d,
it is impossible to find any response function whose predicted light curve
yields an acceptable $\chi^2$, even for $\lambda_* = 0$.
Thus, there is significant flux generated at all
radii from 1 lt-d or less (as argued in the previous paragraph) to at least
8 lt-d (the minimum radius capable of producing a 16d lag).
In other words, the broad emission
line region in NGC 5548 spans at least an order of magnitude
in radius: to speak of a unique ``lag" indicating a unique
radial scale is seriously misleading.  At best, the lag derived from
cross-correlation analysis is no more than a peculiar weighted
average over the response function which is always biassed towards
the smallest lags, with the degree of bias depending on the power
spectrum of continuum fluctuations (\cite{EK88}).
Our finding that the BLR in NGC 5548 spans a large range in radius
illustrates the greater power of true reverberation mapping compared to
cross-correlation analysis.

  Third, the maximal structure solution shows a ``shoulder" roughly
from 8 to 18d lag.  Its absence in the minimal structure solution
demonstrates that this feature, while possibly present, is not required
by the data.  In work which we saw in draft form after submitting
this paper, \cite{W95}, analyzing the same data set with different numerical
methods, independently found a very similar feature, but did not discuss
its statistical
significance.  They interpreted it as evidence for bi-conical structure
in the emission line region.  If the ``shoulder" is real, this is one
possible interpretation, but it is not unique.  It could also be due
to a radial non-uniformity in the emissivity of a perfectly spherical region
(see equation 9).   Which interpretation is preferable is a matter of taste.

  Fourth, the fact that we can find an adequate response function to fit the
lightcurve is in marked contrast to the previous 1989--1990 IUE campaign
(\cite{K91}; \cite{KD95}), although the shape of the derived response
function is qualitatively the same in that it is high at zero lag and
drops to zero by 20 days.  It is entirely possible that the structure
of the broad line region in NGC 5548 has changed significantly over
the four intervening years: the crossing time for material moving at
the median line of sight velocity is only $\simeq 3$yr. Changes in
the H$\beta$ response have also suggested that there is significant
structural change on few year timescales (\cite{bP94}). It is also
possible that there were times during the previous IUE campaign when
the UV continuum was not a good tracer of the total ionizing continuum
flux, perhaps during the third `event' in the continuum lightcurve
where the fractional line change is much larger than in the previous
two `events' (\cite{m94}). For example, a soft X-ray flare was
observed in a ROSAT
monitoring campaign on NGC5548 which was not linearly correlated with
the UV continuum (\cite{D94}; \cite{D95}).

   One problem which first appeared in the 1989 monitoring campaign still
remains, however: the discrepancy between $\partial\ln F_l/\partial \ln
F_{ion}$
(or equivalently $\partial\ln S_l/\partial \ln F_{ion}$)
as estimated from the data and as predicted by photoionization models.  As
we have already remarked, in the likely range of ionization parameters,
it is expected that $\partial\ln F_l/\partial \ln F_{ion} \simeq 1$.  However,
in the earlier experiment, the empirically estimated value was
$\simeq 0.4$ (\cite{K91}).

    Because we are able to find a response function which provides
a good fit to the data for the 1993 experiment, we can estimate the
logarithmic partial derivative by a more reliable method than was
possible using the 1989 data.  This method begins by using
Newton's Theorem applied to equation 8
to solve for the covering factor per unit radius:
\beq
{dC \over dr} \equiv n(r) A(r) = -{4r \over c^2} {d\Psi_{tot} \over d\tau}
\left(\tau=2r/c\right)
{\langle F_{ion} \rangle/S_l(r) \over\partial\ln S_l/\partial \ln F_{ion}}.
\eneq
Since the total line luminosity is
\beq
L_l = \int_{r_{min}}^{r_{max}} \, dr \, 4\pi r^2 {dC \over dr} S_l (r),
\eneq
we may then solve for the emissivity-weighted mean logarithmic
partial derivative:
\beq
\langle {\partial\ln S_l \over \partial \ln F_{ion}} \rangle =
{\langle F_c \rangle \over \langle F_l \rangle}\int_{r_{min}}^{r_{max}} \, dr
 \, {4 r \over c^2}\left[-{d\Psi_{tot}\over d\tau}\left(2r/c\right)\right].
\eneq
Using the maximal structure solution then yields an estimated
$\langle \partial\ln S_l/\partial \ln F_{ion} \rangle \simeq 0.46$,
while the minimal structure
solution gives 0.54.  In other words, just as in the 1989 experiment,
the CIV 1549 line appears to vary significantly more weakly with changing
continuum flux than photoionization models would predict.

   It is possible that the photoionization models are wrong.  However, it
is also possible that the problem is due to this experiment's duration
having artificially imposed a maximum radius which is too small.  In other
words, there could be line-emitting material at such large radius that it
does not respond on the relatively short timescales probed by this
experiment, but does contribute to the time-averaged line flux.  That the
much longer 1989 campaign encountered a similar problem argues against
this interpretation, but that may be a consequence of the fact that no
satisfactory solution for $\Psi_{tot}(\tau)$ could be found from that
campaign's data (\cite{KD95}).

\section{Velocity Resolved Light Curves}

\subsection{Direct comparison of the lightcurve segments}

Figure 5a shows the correlation between the blue core and red core
lightcurves at zero lag.  If both sections of the line respond in the
same way to the continuum fluctuations (i.e. have the same transfer function)
then there should be a linear relationship between them. This is clearly
consistent with the data: a linear regression,
taking into account the fact that the errors on both quantities are of
roughly equal magnitude (\cite{P93}) gives $\chi^2=11.8$ for 37 d.o.f.
Consequently, the velocity structure is consistent with being
completely red/blue symmetric in the core, ruling out pure radial
motions for the BLR.

However, no such linear correlation exists between the blue wing and
red wing lightcurves (Fig. 5b), where $\chi^2=53.4$ for 37 d.o.f.  At
low flux levels, the ratio between the red wing and the blue wing is
much greater than at higher fluxes.  In principle, the contrast between
red and blue wings might be due to contamination of the CIV red wing
by the blue wing of HeII 1640 because the HeII line varies much more
rapidly than does CIV.  The total flux response of HeII is almost
entirely confined within $\simeq 6$d, whereas the CIV 1549 response
must extend to much longer lags (Fig. 4).  We have tested the idea
that HeII contamination accounts for the contrast between the red and
blue wings by computing the red wing--blue wing correlation after
having subtracted a variable fraction $f_{He}$ of the HeII 1640 light
curve from the red wing of CIV 1549.  We find that over the entire
range $0 \leq f_{He} \leq 1$, the smallest $\chi^2$ for the red
wing--blue wing correlation is 44 for 37 degrees of freedom, and that
is achieved for $f_{He} = 0.65$, a magnitude of contamination very
hard to believe.  At the largest plausible $f_{He}$, 20\%, the $\chi^2
= 49$, {\it i.e.}\ there is less than 10\% probability that the red
and blue wings are linearly correlated.

Thus, the picture we have from the raw HST lightcurves alone is
one in which there is a marked similarity between the red and blue
line core lightcurves, indicating that the dominant velocity field is
red/blue symmetric.
Such a symmetric velocity field can be produced
from a disk, or random orbits in 2 or 3 dimensions (\cite{WH91};
\cite{P92}). However, at the same time there is such a strong
contrast between the red and blue wings that there must also be at least
some net radial flow.

\subsection{Response functions}

These inferences from the light curves are substantiated and elaborated
by the response functions.  To illustrate the range of uncertainty due
to choices about our {\it a priori} model, we display in Figs 6a--d
two possible
response functions for each of the four velocity segments: both have
$\tau_{max} = 36$d, but one minimizes deviation from a straight line
and sets $\lambda_* = 1$, while the other minimizes deviation from
a quadratic and sets $\lambda_*$ as large as it can be to still produce
a reduced $\chi^2 \simeq 1$ (this value ranges from $\sim 100$ -- $10^4$
over the four segments).
The former choice produces response functions with
the maximal amount of plausible structure, while the latter
gives response functions with the least structure consistent with
the data. Again, the line and continuum lightcurves
have been sampled 100 times within the measurement error distribution,
and the error bars in Figs 6a--d show the effect of the observational
uncertainty on the minimal emissivity structure model.
Error bars for the maximal emissivity structure model are supressed
for clarity, but they are typically larger due to the smaller value
of $\lambda*$. Again we stress that these error bars are not independent,
but are strongly correlated from point to point.

\subsection{Qualitative interpretation}

   Independent of the detailed choices made for the solutions, two points
stand out from these curves: the red and blue cores are nearly identical
in their response; and the red wing response has a strong
peak at small lags, but the blue wing response rises only very slightly
towards $\tau = 0$.

The similarity between the red and blue cores at all lags suggests that,
to zeroth order, there is
little correlation between the local mean radial velocity of the
line-emitting gas in NGC 5548 and position.  In addition, the weak
dependence of the core/wing ratio on lag suggests that the magnitude of the
mean velocity depends only weakly on radius.

   At the same time, however, the
dramatic contrast between the blue and red wings at small lag indicates
that there is more material near the line of sight and/or at small
radii which is travelling away from us than is travelling towards us.
This suggests that any net radial flow is toward the central object.

Finally, in all the maximal structure responses, there is a sharp minimum
(which in some cases actually reaches negative values) between 20 and 25d
and a ``shoulder" at 10 to 15d.  Both features are also seen in the total
flux response function.  As we discussed in \S 4, while these features are
plausible, they are not required by the data.  Just as for the total
flux response function, similar features were also found, but not tested
for statistical significance, by \cite{W95}.

\section{Modelling}

    Unfortunately, more quantitative statements are difficult to make solely
on the basis of these response functions.  In order to clarify our
conclusions, we have embarked on a program of direct modelling.  While
the forms that our models take are guided qualitatively by our solutions
for the velocity-resolved response functions, the parameter fits we perform
are, with one exception (the emissivity as a function of radius---see
the discussion below), statistically quite independent of these prior
solutions.

\subsection{Model construction}

Equation 7 gives the general form for the velocity-resolved response function.
The response function corresponding to one of our four segments
is simply $\Psi (\tau,u)$ integrated over the appropriate range of
$u$.  In principle it would take a very large number of parameters to describe
all the functions determining the four response functions, but for obvious
reasons it is desirable to find models with the least number of free parameters
consistent with arriving at an acceptable description of the data.

Fortunately, although the cloud numbers, distribution and response are
all unknown, it is not necessary to define each of them explicitly, or
even to guess a model for them.  Instead, guided by the shape of the total flux
response function, we make the {\it ansatz} that the CIV line emissivity is a
function of radius alone (spherical symmetry), and that the material emits
isotropically.  As we have previously demonstrated, if these assumptions apply,
the marginal emissivity as a function of radius may be simply read off the
total flux response function  $\Psi_{tot}(\tau)$ (see equation 9).

As we have also already discussed, it is hard to choose a form for the response
function which is uniquely best.  To illustrate the effects of this
range of uncertainty, we have done model fits deriving the emissivity
as a function of radius from two response functions we believe to span
the likely range.  These are the ``minimal structure" response function
($\tau_{max} = 36$d, $\lambda_* = 10^4$, and a quadratic regularization
condition, Fig. 3), and the maximal plausible
structure response function ($\tau_{max} = 36$d, $\lambda_* = 1$, and a
linear regularization condition, solid line in Fig. 4).

To complete the model specification all we need to do is construct a
parameterized form for the velocity distribution function as a function of
$r$.  We investigate two forms for this distribution, one combining
radial motion with 3-d random motions, the other combining radial motion
with random motions confined to the local tangent plane.  By this means
we test whether there is a preferred eccentricity for the
random orbital motions.

The velocity distribution function for the first form may be
written as:
$f[{\bf v} - v_r (r){\bf \hat r}]=\sigma(r)^{-3}\pi^{-3/2}
\exp[-(v_x^2+v_y^2+v_z^2)/\sigma(r)^2]$.
That is, in this form we suppose that the velocity distribution at each point
is centered on radial velocity $v_r (r) {\bf \hat r}$, but with a Gaussian
spread
characterized by $\sigma(r)$.  With this definition, we have
\beq
\Psi_l (\tau,u)=\int_{c\tau/2}^{r_{max}=c\tau_{max}/2} \, dr \,
\left( -{d\Psi_{tot}(\tau)\over d\tau} \right){2\over c}
{e^{-[u-v(r)(1-c\tau/r)]^2/\sigma(r)^2} \over \sqrt{\pi} \sigma(r)} .
\eneq
The last step in specifying such a ``random + radial" model is to
define $v_r(r)$ and $\sigma(r)$.  We choose the forms
\beq
v_r(r) = v_o (r/r_o)^{\alpha}
\eneq
and
\beq
\sigma(r) = \sigma_o (r/r_o)^{\beta},
\eneq
where $r_o$ is a fiducial radial scale, which we set at 1 lt-d.  Thus, this
class of model is determined completely by only four free parameters.

Our other form restricts the random motions to the local tangent plane.
In this case, rather than defining the velocity distribution in terms of
its density in 3-d velocity space, it is most simply described
in terms of the total magnitude of the velocity and the direction of
the component in the tangent plane.  That is, if we define a local set of
basis vectors in the tangent plane:
\begin{eqnarray}
{\bf \hat x_t} &= -\sin\theta {\bf \hat z} + \cos\theta \cos\phi {\bf \hat x}
+ \cos\theta \sin\phi {\bf \hat y} \\
{\bf \hat y_t} &= -\sin\phi {\bf \hat x} + \cos\phi {\bf \hat y} ,
\end{eqnarray}
we may write the tangential velocity as
${\bf v_t} = v_t(r)(\cos\gamma {\bf \hat x_t} + \sin\gamma {\bf \hat y_t})$.
Then the 2-d random velocity distribution is simply
\beq
df(v,\gamma) = \delta \left(v - \sqrt{v_r^2 + v_t^2}\right)
{dv \, d\gamma \over 2\pi}.
\eneq
With this form for the velocity distribution, the response function becomes
\beq
\Psi (\tau,u) = {1 \over \pi c} \int_{r_1}^{r_2} \, dr \,
\left( -{d\Psi_{tot}(\tau)\over d\tau}\right)
 \left\{ v_t^2 \left[2{c\tau \over r} - \left({c\tau \over r}\right)^2\right]
- \left[ v_r \left( 1 - {c\tau \over r}\right) - u\right]^2 \right\}^{-1/2}.
\eneq
where $r_1$ and $r_2$ are the radii limits over which the integrand exists.

Once again, the most concise parameterization of the characteristic
velocities is as power-laws in radius.  We retain the same notation for
$v_r (r)$; we write $v_t (r)$ as
\beq
v_t (r) = v_{to} (r/r_o)^{\epsilon}.
\eneq

We integrate these equations numerically using an evenly spaced grid
of 200 points in velocity space, and oversample the $\tau$ points by a
factor 10, linearly interpolating for the values of $d\Psi/d\tau$
within this. For the 2--d random motions, the nature of the integrand is such
that a smooth response function can be better obtained by averaging over the
oversampled points either side of the desired $\tau$.

\subsection{Model Fits}

The most natural way to measure the quality of a model is, of course,
to compute its $\chi^2$ with respect to the data and compare that
$\chi^2$ to the number of degrees of freedom present.  In this application,
while it is obvious how to compute $\chi^2$ (generate a response function
from the model, fold it through the measured continuum light curve, and
compare the results to the observed line light curves in each of the
four segments), it is not quite so obvious how to compute the number of
degrees of freedom.  In each of the four line light curves smoothed to
2d resolution, there are 20 points.  However, we are fitting to their
fluctuations about the mean, so they are constrained to sum to zero.
This reduces the number of degrees of freedom by one for each light curve.
In addition, we have removed the mean from the continuum light curve,
so on this basis one might expect that there are
$4 \times (20 - 1) - 1 -n_{par} = 75 - n_{par}$
degrees of freedom, where $n_{par}$ is the number of parameters in
the model.  However, the expected error in the predicted line
light curve must also include the error induced by noise in the continuum
light curve.  To account for this effect, we increase the effective
number of degrees of freedom in each line light curve by the same factor
dependent on the comparative S/N of the continuum and line light curves
that we used to judge the quality of the regularized inversions (see
\cite{KD95}).  These ratios are 1.08, 1.10, 1.12, and 1.14, for the
blue wing, blue core, red core, and red wing, respectively.  On
this basis, the
total number of effective degrees of freedom becomes $83.36 - n_{par}$.

Unfortunately, $n_{par}$ is not altogether unambiguous, either.  In each model
for the velocity distribution examined below, there are (depending on
the model) between 1 and 4 free parameters.  {\it If one were to
assume that the radial emissivity derived from our fit to the total
line flux lightcurve is correct}, $n_{par}$ would include only those.
However, another interpretation would
be to include the radial emissivity as part of the model being tested.  In
that case, one would count an additional 2 free parameters when the
minimal structure emissivity is used (a parabolic fit with
$\Psi_{tot}(\tau_{max})$ fixed at zero), or possibly as many as 19
(the total number of elements in the solution vector) in the limit of
$\lambda_* \rightarrow 0$! Simulations show that $\lambda*=1$
lies approximately half way between these two extremes (\cite{KD95})
in terms of the effective number of degrees of freedom,
so that there are an additional $\sim 10$ free parameters
from the radial emissivity in the maximal structure solution.
However, since this latter number is rather poorly defined, and so
as to have a fixed standard of comparison,
we will adopt the definition that the radial emissivity introduces 2
additional free parameters, but we stress that {\it changes} in
reduced $\chi^2$ are more important for our argument than the absolute
value of the reduced $\chi^2$.  With that definition, the models have between
77.4 and 80.4 effective degrees of freedom.

We begin by demonstrating that the three simplest models can be
easily ruled out, no matter how the free parameters are counted (the data
for the models we describe here and in subsequent paragraphs are summarized
in Tables 2 and 3.)
If we fix $\sigma_o = \beta = 0$ ({\it i.e.\ } a
model with purely radial motion), then the best fit model (using
either form for the total flux response) has a reduced
$\chi^2 \simeq 7$!  Less trivially, a purely random model can also be
eliminated.  With $v_o = \alpha = 0$, the best fit
has a reduced $\chi^2 = 2.20$ for the ``minimal structure" response,
or 1.88 for the ``maximal structure" form.  The fit qualities for
the best purely tangential models are very similar to the best
purely random models.  Making any allowance for additional implicit free
parameters in the maximal structure solution would only {\it increase}
the least reduced $\chi^2$.

However, as we have already inferred from the qualitative character
of the velocity-resolved response functions, a combination of either
3-d or 2-d random motion plus inward radial flow does much better.
Combined 3-d random and radial velocity fields give a
best reduced $\chi^2 \simeq 1.34$ for the maximal structure
emissivity when $v_{o} \simeq -790$ km s$^{-1}$, $\alpha \simeq 0.35$,
$\sigma_o \simeq 5600$ km s$^{-1}$, and $\beta = -0.14$.
The four light curves predicted by this
model are compared to the observed light curves in Figs. 7.

There is no simple description for the uncertainty ranges of the
fit parameters because there are strong mutual correlations between
the parameters, and their
acceptable ranges depend on the form of the model.  In order
to give some sense of this uncertainty, we discuss the effects of
altering various model assumptions in the following paragraphs.

The formal reduced $\chi^2$ of the combined random + radial model
depends on the description of the emissivity: using the minimal
structure emissivity increases the least reduced $\chi^2$ to 1.71.
Most of this increase can probably be attributed to the greater
$\chi^2$ of the total flux response function which defines the
emissivity in this model.
In addition, the minimal structure emissivity leads to rather different
(and somewhat surprising)
best-fit parameters: $v_{o} \simeq -0.04$ \kms, $\alpha \simeq 4.3$,
$\sigma_o \simeq 22,000$ \kms, and $\beta \simeq -0.8$.  Nonetheless, the
{\it reduction} in reduced $\chi^2$ due to adding radial motions to
the model is nearly independent of the emissivity model: in the maximal
structure case, reduced $\chi^2$ falls by 0.54, while in the minimal
structure case, it is diminished by 0.49.  We take this to indicate
that a combination of radial and random motion is required for
any total flux emissivity consistent with the data.

2-d and 3-d random motions are indistinguishable in this regard.  With
the maximal structure emissivity, the least 2-d reduced $\chi^2$
for a random + radial model is 1.37, and this is
achieved for parameters very similar to those found in the 3-d solution:
$v_{o} \simeq -820$ km s$^{-1}$, $\alpha \simeq 0.35$,
$v_{to} \simeq 5200$ km s$^{-1}$, and $\epsilon = -0.13$.  2-d motions without
any radial velocity lead to a reduced $\chi^2$ larger by $\simeq 0.9$.
We therefore conclude that these data do not discriminate between orbits of
differing eccentricity.

   The magnitude of the random velocity in the middle of the emission region
is nearly model-independent; at 5 lt-d it is between 4200 and 5400 \kms\ in
essentially all the random + radial models.  This is, of course, because
they must match the characteristic width of the velocity profile.  Similarly,
the magnitude of the radial velocity at 5 lt-d is in the range $-1400$ to
$-2000$ \kms\ for all the successful maximal structure models, although
significant radial velocities tend to be reached only at larger radii in
the minimal structure models.

The degree to which we may constrain the magnitudes of the velocity
gradients ($\alpha$, $\beta$, and $\epsilon$) can depend on one's
choice of model as well as the choice of emissivity.  For example,
while fixing $\beta = -0.5$ in the maximal structure 3-d random + radial
model raises reduced $\chi^2$ by $\simeq 0.25$, the increase in reduced
$\chi^2$ when $\epsilon$ is set to -0.5 in the maximal structure
2-d random + radial model is only 0.08, while the same exercise applied
to the minimal structure models increases $\chi^2$ by even smaller amounts.

   Not surprisingly, it is even harder to constrain gradients in the radial
velocity.  In minimal structure models, the best-fit values of $\alpha$
are generally 3 -- 4; forcing $\alpha$ to be as small as 0 increases reduced
$\chi^2$ by $\simeq 0.1$ -- 0.2.  That is, with this description of the
emissivity, $\alpha$ is likely to be fairly large and positive, but smaller
gradients are not strongly ruled out.  On the other hand, in the maximal
structure models, the best-fit values are $\simeq 0.3$ -- 0.4, and
forcing $\alpha = 0$ has only a trivial effect on $\chi^2$ in the 3-d
case, but can be ruled out if the motions are 2-d.  Thus, while the
best-fit value of $\alpha$ is generally positive, it has a large range of
uncertainty.

   We conclude by pointing out the one possible escape from our conclusion
that both radial infall and random motions must be present in the BLR
of NGC 5548: the difficulty all models have in fitting the red wing.
Because the red wing lightcurve consistently gives a worse fit than any
of the other velocity segments, one might suspect that there is
an unidentified problem in the red wing data.  We have already argued
against any significant contamination from HeII 1640, but there may
be some other problem which disproportionately affects this segment of
the line.  To test this hypothesis, we have also examined
models fit to the three other line segments, but ignoring the red wing.
Dropping the red wing does improve the fit: a maximal structure
emissivity random + radial model with $\alpha = \beta = 0$ then gives
$\chi^2=1.07$ for $v_o=-1500$ km s$^{-1}$ and $\sigma_o=4200$ km s$^{-1}$;
similarly, without the red wing the minimal structure emissivity permits
an analogous model with reduced $\chi^2 = 1.41$.
However, even without the red wing, the maximal structure emissivity
still requires radial motion: if we force $v_o = 0$
while ignoring the red wing, the least reduced $\chi^2$ for
the maximal structure emissivity increases by 0.2 even allowing $\beta$
to vary.  Only if one restricts attention to the minimal structure emissivity
model and ignores the red wing does the omission of radial motion not
significantly increase $\chi^2$.

     Comparing the best-fit model red wing response with the curve derived
directly from the data suggests a more likely explanation for our
problems in fitting the red wing: the relatively short duration of
the experiment.  Adopting our {\it ansatz} of spherically symmetric,
isotropically radiated emission means that the largest radius whose
emissivity is revealed by the total flux response function is 18 lt-d;
our models therefore have {\it zero} response at larger radii.  When
there is some radial infall, however, the lag in the red wing at
$\tau > 20$d depends largely on material at radii $> 20$ lt-d.  Consequently,
the absence of data on longer lags cripples the ability of our models
to account for the red wing's response beyond $\simeq 20$d.

\section{Discussion: Comparison with Previous Work, and
Once Popular Models Now Ruled Out}

    On the basis of the arguments presented in the previous sections, we have
now arrived at a clearly-established qualitative picture of the kinematics
in the BLR of NGC 5548:  Significant CIV 1549 emission stretches over
at least a factor of ten in radius, from 1 lt-d or closer, out to 10 lt-d
or farther.  At any particular location within that region, the
velocity of the line-emitting gas exhibits a fairly broad distribution,
but there is a net tendency toward inward motion.  The magnitude of
the mean inward speed is perhaps a few times smaller than the characteristic
spread in velocities, and that characteristic spread may vary rather slowly
with radius, but is not absolutely required to do so.

Elements of this picture had been hinted at in several previous
studies.  Suggestive evidence for a combination of random motion and
radial infall has been noted by \cite{KG89} in the case of Fairall 9,
and by \cite{M91} in NGC 4151.  \cite{WH94} advocated predominantly
random motion in NGC 3516, but commented that the contrast between core
and wing predicted by Keplerian motions was not present (although they made
no specific test to see whether velocities scaling as $r^{-0.5}$
were inconsistent with the data).  While previous
studies of NGC 5548 have produced contradictory results, some suggested
pieces of what we find: \cite{KG91a} contended (on the basis of data
with a mean sampling interal of 97d) that the motion
is predominantly random, while \cite{CB90} argued for strong radial infall.
The superior
quality of the data provided by the {\it HST} monitoring experiment has
allowed for the first time a truly quantitative approach to this
problem, so that the relative contributions of random and
radial motion can at last be quantitatively assessed.

   Although significant error bars remain for many of the interesting
parameters, the data in hand already now allow us to rule out several of the
historically most popular models for the broad line region's dynamics.

   Almost all wind models, whether propelled by radiation pressure
(\cite{M74}, \cite{BM75}; \cite{M86}), thermal winds
(\cite{W82}; \cite{B90}), or rotating magnetic fields (\cite{E92}) can
be immediately discarded because they predict
net outflow, not net infall.  Those wind models invoking
a fluid substrate are ruled out with special force because they would
also find it hard to accommodate significant random velocities.  The only
exception to this conclusion is that class of models in which obscuration
prevents us from seeing the far side of the source.  Because the center
of the line profile is now made by material that is actually moving
towards us, the simple identification of red and blue with near and far sides
which allowed us to so easily rule out simple wind models is destroyed.
However, these models are
still in contradiction with the data if, as they frequently do ({\it e.g. }
\cite{M95}), they predict
the greatest speeds to occur at the smallest radii.  This is because in
that case the blue wing would still tend to respond at the smallest lags.

Randomly oriented gravitational orbits, whether of clouds
(\cite{KC82}) or stars (\cite{P88}; \cite{NS88}; \cite{K89};
\cite{AN94}), and having any distribution of eccentricities, are
also strongly ruled out by the fact that there is net radial
infall.  The model of \cite{CK85}, in which randomly oriented orbits
are combined with drag against an external wind is still viable.  In
this model one does expect a combination of random velocities and net
infall.

Emission from the surface of accretion disks (\cite{JR80}; \cite{GP81};
\cite{M82}; \cite{vG83}) is ruled out by the same
arguments which eliminate randomly oriented orbits.  They would
also be subject to the requirement that the disk must be highly inclined
(close to edge on) in order for the total flux response to
peak at zero lag, which is in conflict with the detection of a
strong Compton reflection spectrum in the X--ray spectrum
(\cite{NP94}). The velocity field from a disk is also
red/blue symmetric (\cite{WH91}; \cite{P92}),
and so a disk geometry alone cannot explain the
significant differences between the blue and red wing lightcurves.
Some sort of combination of a disk with infalling gas would be
required in order to begin to match the kinematics.

Smooth radial accretion (\cite{KL83}) also cannot
explain these data.  Models of this variety cannot allow substantial
random velocities.

The fundamental reason why these data are in conflict with all simple
theoretical models is that the real situation appears to be an intermediate
case: not entirely collisionless, but definitely not fluid-like.  No
one would begin by proposing such a complicated hybrid.  However, the
behavior of line variations in this experiment appears to require just
such a model.  While the single largest component of the motion is
random (whether fully 3-d or restricted to the tangent plane), there
also seems to be a secondary component of radial infall whose origin
must presumably lie in some sort of dissipative process.

  We close with one final comment.  One frequently finds in the
literature attempts to estimate the central mass in AGN by multiplying
a ``characteristic'' radius by the square of a ``characteristic''
velocity.  The analysis we have presented in this paper demonstrates
clearly that there is at least an order of magnitude dynamic range
between the inner and outer radii bounding the zone of substantial
broad line emission in NGC 5548.  This fact alone calls into question
these simple estimates.  If, in addition, the hints we have seen that
$\beta$ or $\epsilon$ are significantly different from -0.5 are confirmed,
the whole physical basis of the exercise would be severely undermined.

    A better procedure would be to work as follows:  First,
construct a model for the emissivity as a function of
position in which the orbital speeds are consistent with the assumed
shape of the potential, and check whether this model provides an
adequate description of the data.  For example, the maximal structure
emissivity with $\beta$ or $\epsilon$ set to -0.5 yields a not completely
satisfactory reduced $\chi^2 \simeq 1.5$ (the best-fit $\sigma_o$
and $v_t$ using the minimal structure emissivity are the same as
for the maximal structure emissivity, but give larger $\chi^2$).
Second, relate the velocities
and the central mass through the physical model.  In the 3-d + random model,
this would mean that $M = 3 r_o v_o^2/G$, while in the 2-d + random
model $M = r_o v_o^2/G$.  Thus, with Keplerian orbital dynamics
forced on the data,
the central mass might be either $8 \times 10^7 M_{\odot}$ (for 3-d
random motions) or $2\times 10^7 M_{\odot}$ (for 2-d random motions).
Given the large uncertainty in both procedures, it is quite surprising that
these estimates are almost identical to those made in
\cite{K91} on the basis of the correlation between profile width and
characteristic response time for eight different emission lines in the
spectrum of NGC 5548.   The only point of contact between the two
estimates is that both must be consistent with the characteristic velocity
width and response time for CIV 1549 in this object.  Despite this
striking agreement, in view of the
crudeness of the quality of fit in both procedures, and the possible
inapplicability of the Keplerian model, we remain
reluctant to ascribe too much reality to these estimates of the central mass.

\acknowledgments

This work was partially supported by NASA Grant NAGW-3156.
CD is funded by a PPARC Advanced Fellowship. We thank Keith Horne and
Bill Welsh for some illuminating discussions about velocity-resolved
transfer functions, and Richard Mushotzky for comments on non--linear
effects in the line.  We also thank Ignaz Wanders for sending us a
draft of his group's work on the same topic.

\begin{deluxetable}{cccccc}
\tablecolumns{6}
\tablewidth{0pt}
\tablecaption{Velocity resolved CIV line lightcurves from HST
with 2\% systematic errors}
\tablehead{
\colhead{line} &
\colhead{$<F>$} &
\colhead{$\sigma^2_{tot}$} &
\colhead{$\sigma_{err}^2$} &
\colhead{$\delta F/<F>$} &
\colhead{S/N}}
\startdata
total continuum & $3.49\times 10^{-14}$ & $2.61\times 10^{-29}$ &
$8.40\times 10^{-31}$ & 0.144 & 5.48\nl
total line & $6.53\times 10^{-12}$ & $3.75\times 10^{-25}$ &
$1.78\times 10^{-26}$ & $0.091$ & 4.48\nl
bw & $1.37\times 10^{-12}$ & $1.37\times 10^{-26}$ & $9.05\times 10^{-28}$ &
0.083  & 3.76 \nl
bc & $1.83\times 10^{-12}$ & $2.74\times 10^{-26}$ & $1.48 \times 10^{-27}$ &
0.097\tablenotemark{a}  & 4.23 \nl
rc & $1.96\times 10^{-12}$ & $3.54\times 10^{-26}$ & $1.67\times 10^{-27}$ &
0.10\tablenotemark{a} & 4.52 \nl
rw & $1.37\times 10^{-12}$ & $2.27\times 10^{-26}$ & $9.24\times 10^{-28}$ &
0.11  & 4.85 \nl
\enddata
\tablenotetext{a}{Calculated allowing for a 10\% contribution to the mean
flux from the narrow line component.}
\tablecomments{All fluxes are in ergs s$^{-1}$ cm$^{-2}$ except for the
continuum for which the units are ergs s$^{-1}$ cm$^{-2}$ A$^{-1}$.
All variances are in ergs$^2$ s$^{-2}$ cm$^{-4}$ except
for the continuum for which the units are
ergs$^2$ s$^{-2}$ cm$^{-4}$ A$^{-2}$.}
\end{deluxetable}

\begin{deluxetable}{cccccccccc}
\tablecolumns{10}
\tablewidth{0pt}
\tablecaption{Fitting the velocity resolved CIV line segments: minimal
structure}
\tablehead{
\colhead{Model} &
\colhead{$v_o$ km $s^{-1}$} &
\colhead{$\alpha$} &
\colhead{$\sigma_o$\tablenotemark{a},\ $v_t$\tablenotemark{b}} &
\colhead{$\beta$\tablenotemark{a},\ $\epsilon$\tablenotemark{b}} &
\colhead{$\chi^2$(bw)} &
\colhead{$\chi^2$(bc)} &
\colhead{$\chi^2$(rc)} &
\colhead{$\chi^2$(rw)} &
\colhead{$\chi^2$(Total)}
}
\startdata
R & $-4700$ & 0\tablenotemark{c} & & & 171.4 & 60.2 & 58.1 & 252.3 & 542.1\nl
R & $-2100$ & 0.42 & & & 231.6 & 63.2 & 58.3 & 175.8 & 529.0\nl
3D & & & $4500$ & 0\tablenotemark{c} & 36.5 & 24.3 & 29.3 & 93.6 & 183.6 \nl
3D & & & $11,000$ & -0.44 & 44.8 & 22.6 & 31.6 & 76.0 & 174.9 \nl
2D & & & $4100$ & 0\tablenotemark{c} & 42.9 & 21.2 & 29.2 & 86.9 & 180.2 \nl
2D & & & $5600$ & -0.15 & 43.7 & 18.8 & 30.3 & 79.3 & 172.0\nl
3D+R & $-1100$ & 0\tablenotemark{c} & $4400$ & 0\tablenotemark{c} & 39.8 & 23.8
& 30.3 & 75.3 & 169.2\nl
3D+R & $-1100$ & 0\tablenotemark{c} & $13,000$ & -0.54 & 39.6 & 23.5 &
30.1 & 64.7 & 157.9\nl
3D+R & $-0.4$ & 3.4 & $4400$ & 0\tablenotemark{c} & 35.8 & 22.2 & 29.2 & 61.0 &
148.2\nl
3D+R & $-0.04$ & 4.30 & $22,000$ & -0.81 & 33.7 & 22.2 & 27.6 &
48.7 & 132.2 \nl
3D+R & $-0.6$ & 3.31 & $12,000$ & -0.50\tablenotemark{c} & 33.1 & 21.7 &
27.8 & 51.3 & 133.9\nl
2D+R & $-570$ & 0\tablenotemark{c} & $4100$ & 0\tablenotemark{c} & 29.7 & 20.9
&
28.5 & 77.0 & 156.1\nl
2D+R & $-820$ & 0\tablenotemark{c} & $5500$ & -0.16 & 31.3 & 20.0 &
27.4 & 75.2 & 153.9\nl
2D+R & $-3.4$ & 2.53 & $4000$ & 0\tablenotemark{c} & 32.3 &  20.4 &
27.9 & 61.9 & 142.5\nl
2D+R & $-0.4$ & 3.23 & $8000$ & -0.33 & 29.5 & 18.3 & 29.6 & 49.0 & 126.5 \nl
2D+R & $-0.5$ & 3.06 & $12,000$ & -0.50\tablenotemark{c} & 38.3 & 18.5 &
25.2 & 51.8 & 133.8 \nl
\enddata
\tablenotetext{a}{3--d random motions}
\tablenotetext{b}{2--d random motions}
\tablenotetext{c}{Parameter fixed}
\tablecomments{The model coding is: R for purely radial; 3D for 3-d random
motions; 2D for 2-d random motions.  The effective total number of degrees
of freedom is $\simeq 77$ -- 80.}
\end{deluxetable}

\begin{deluxetable}{cccccccccc}
\tablecolumns{10}
\tablewidth{0pt}
\tablecaption{Fitting the velocity resolved CIV line segments: maximal
structure}
\tablehead{
\colhead{Model} &
\colhead{$v_o$ km $s^{-1}$} &
\colhead{$\alpha$} &
\colhead{$\sigma_o$\tablenotemark{a},\ $v_t$\tablenotemark{b}} &
\colhead{$\beta$\tablenotemark{a},\ $\epsilon$\tablenotemark{b}} &
\colhead{$\chi^2$(bw)} &
\colhead{$\chi^2$(bc)} &
\colhead{$\chi^2$(rc)} &
\colhead{$\chi^2$(rw)} &
\colhead{$\chi^2$(Total)}
}
\startdata
R & $-5100$ & 0\tablenotemark{c} & & & 86.8 & 261.8 & 62.1 & 248.0 & 658.7\nl
R & $-1900$ & 0.52 & & & 172.5 & 135.1 & 88.1 & 178.6 & 574.2\nl
3D & & & $4500$ & 0\tablenotemark{c} & 39.9 & 22.0 & 20.2 & 71.1 & 153.3\nl
3D & & & $6100$ & -0.16 & 43.9 & 20.3 & 20.1 & 65.1 & 149.4\nl
2D & & & $4200$ & 0\tablenotemark{c} & 48.6 & 29.8 & 32.8 & 90.5 & 201.7 \nl
2D & & & $5300$ & -0.12 & 40.5 & 27.5 & 25.1 & 81.1 & 174.3\nl
3D+R & $-2000$ & 0\tablenotemark{c} & $4200$ & 0\tablenotemark{c} & 24.8 & 19.2
& 20.6 & 46.1 & 110.7\nl
3D+R & $-1800$ & 0\tablenotemark{c} & $5600$ & -0.14 &
23.0 & 18.4 & 18.9 & 47.2 & 107.5\nl
3D+R & $-910$ & 0.33 & $4300$ & 0\tablenotemark{c} &
24.5 & 18.1 & 19.8 & 44.4 & 106.9\nl
3D+R & $-790$ & 0.35 & $5600$ & -0.13 &
23.7 & 17.3 & 18.2 & 44.6 & 103.8\nl
3D+R & $-390$ & 0.50 & $12,000$ & -0.50\tablenotemark{c} &
33.7 & 23.0 & 18.5 & 47.6 & 122.9\nl
2D+R & $-700$ & 0\tablenotemark{c} & $4000$ & 0\tablenotemark{c} &
37.7 & 33.1 & 24.6 & 56.6 & 152.1\nl
2D+R & $-750$ & 0\tablenotemark{c} & $5200$ & -0.13 &
25.8 & 25.2 & 18.3 & 59.8 & 130.0\nl
2D+R & $-830$ & 0.38 & $3900$ & 0\tablenotemark{c} &
38.3 & 18.9 & 18.8 & 45.1 & 121.2\nl
2D+R & $-820$ & 0.35 & $5200$ & -0.13 &
25.5 & 16.0 & 18.3 & 46.3 & 106.0\nl
2D+R & $-81$ & 0.79 & $12,000$ & -0.50 &
27.4 & 18.4 & 19.3 & 47.3 & 112.4\nl

\enddata
\tablenotetext{a}{3--d random motions}
\tablenotetext{b}{2--d random motions}
\tablenotetext{c}{Parameter fixed}
\tablecomments{The model coding is: R for purely radial; 3D for 3-d random
motions; 2D for 2-d random motions.  The effective total number of degrees
of freedom is $\simeq 77$ -- 80.}
\end{deluxetable}

\clearpage

\centerline{Figure Captions}

\vskip 0.4cm

\noindent Figure 1 \qquad The CIV total flux response function making
use of only {\it HST} data, setting $\tau_{max} = 18$d, and $\lambda_* =
10^4$. The error bars show the uncertainty in the derived response
due to the measurement errors and are correlated from point to point.
These errors  do not include the inherent
non--uniqueness in the deconvolution due to the particular
parameter choices of $\tau_{max}$ and $\lambda_*$.

\noindent Figure 2 \qquad The CIV total flux response function making
use of only {\it HST} data, setting $\tau_{max} = 12$d.  The
smooth dotted curve shows the result of setting $\lambda_* = 750$, the largest
value of $\lambda_*$ for which reduced $\chi^2 \leq 1$; the jagged solid
curve with error bars (derived as for Fig. 1)
is the solution for $\lambda _* = 1$.
The larger
response at $\tau = 0$ compared to Fig. 1
likely reflects the influence of the systematic
calibration errors.

\noindent Figure 3 \qquad The smoothest possible CIV total flux
response function using the combined data set for the continuum light
curve.  Here $\lambda_* = 10^4$, but the solution is virtually independent
of $\lambda_*$ for any value greater than $\sim 100$. Error bars
are derived as in Fig. 1.

\noindent Figure 4 \qquad The combined {\it IUE} + {\it HST} continuum
light curve
permits acceptable solutions with $\tau_{max} \geq 16$d.  All the
response function solutions shown here used $\lambda_* = 1$ and have
reduced $\chi^2 \leq 1$.  Error bars are suppressed for clarity.

\noindent Figure 5a \qquad The correlation between the blue core and red core
lightcurves.  The fluxes are in units of $10^{-12}$ erg cm$^{-2}$ s$^{-1}$.
The ratio between the blue core's fluctuation about its
mean and the red core's fluctuation about its mean is clearly constant.
However, the extrapolation of this relation to zero flux in
both segments does require some curvature. The best fit linear
correlation (not constrained to pass through zero)
has $\chi^2=12$ for 37 degrees of freedom.

\noindent Figure 5b \qquad The correlation between the blue wing and
red wing lightcurves.  The fluxes are in units of $10^{-12}$ erg cm$^{-2}$
s$^{-1}$.  Their fluctuations are clearly inconsistent with any
simple linear relation, as evinced by the best fit $\chi^2$ of
53.4 for 37 degrees for freedom.

\noindent Figure 6a \qquad The response function for the red wing.  In
this and the other three panels of Fig. 6, the solid line with error
bars (derived as in Fig. 1)
is the minimal structure response function, and the dotted line is the
maximal structure response function.  Error bars are shown only for the former
solution for clarity.

Note that this response function has large amplitude at small lag, in sharp
contrast to the blue wing response function (Fig. 6d).

\noindent Figure 6b \qquad The response function for the red core.  It
is consistent with being identical to the blue core.

\noindent Figure 6c \qquad The response function for the blue core.

\noindent Figure 6d \qquad The response function for the blue wing.

\noindent Figure 7a -- d \qquad Predicted vs. observed lightcurves for:
(a) the red wing; (b) the red core; (c) the blue core; and (d) the blue wing
in the best fit combined radial and random motion model.  In all four
panels the solid line shows the predicted lightcurve and the points with
error bars are the observations.

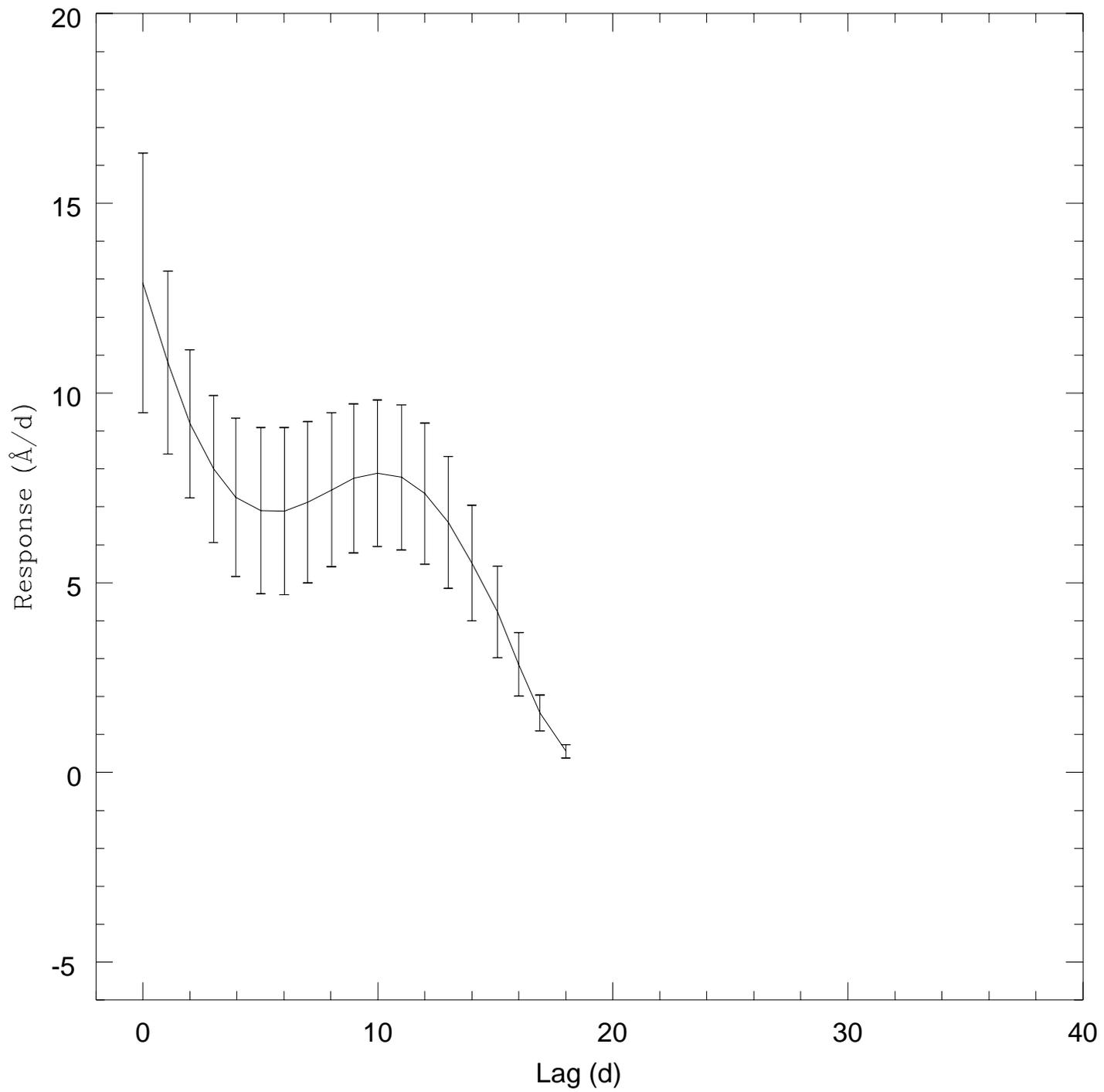
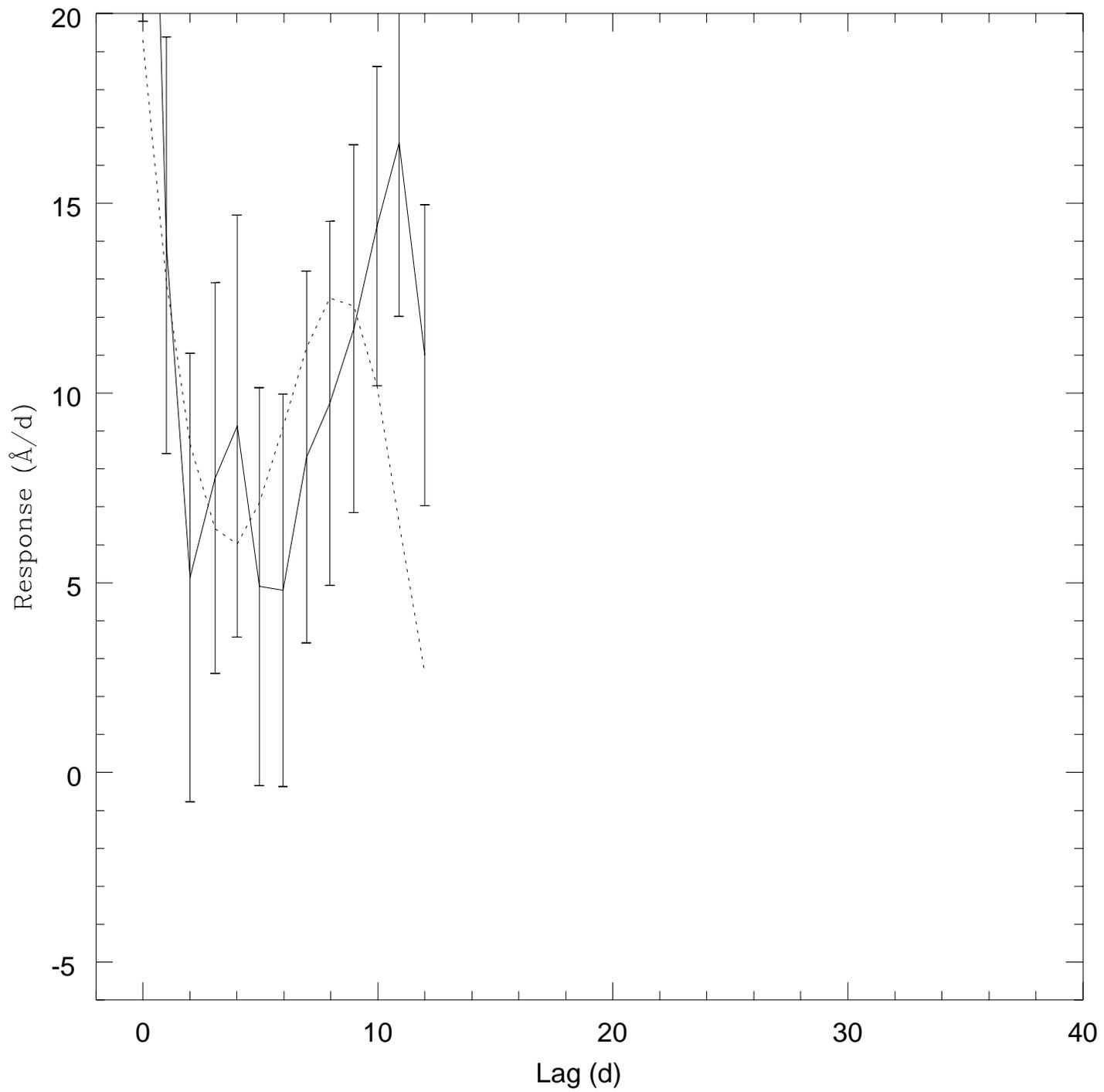
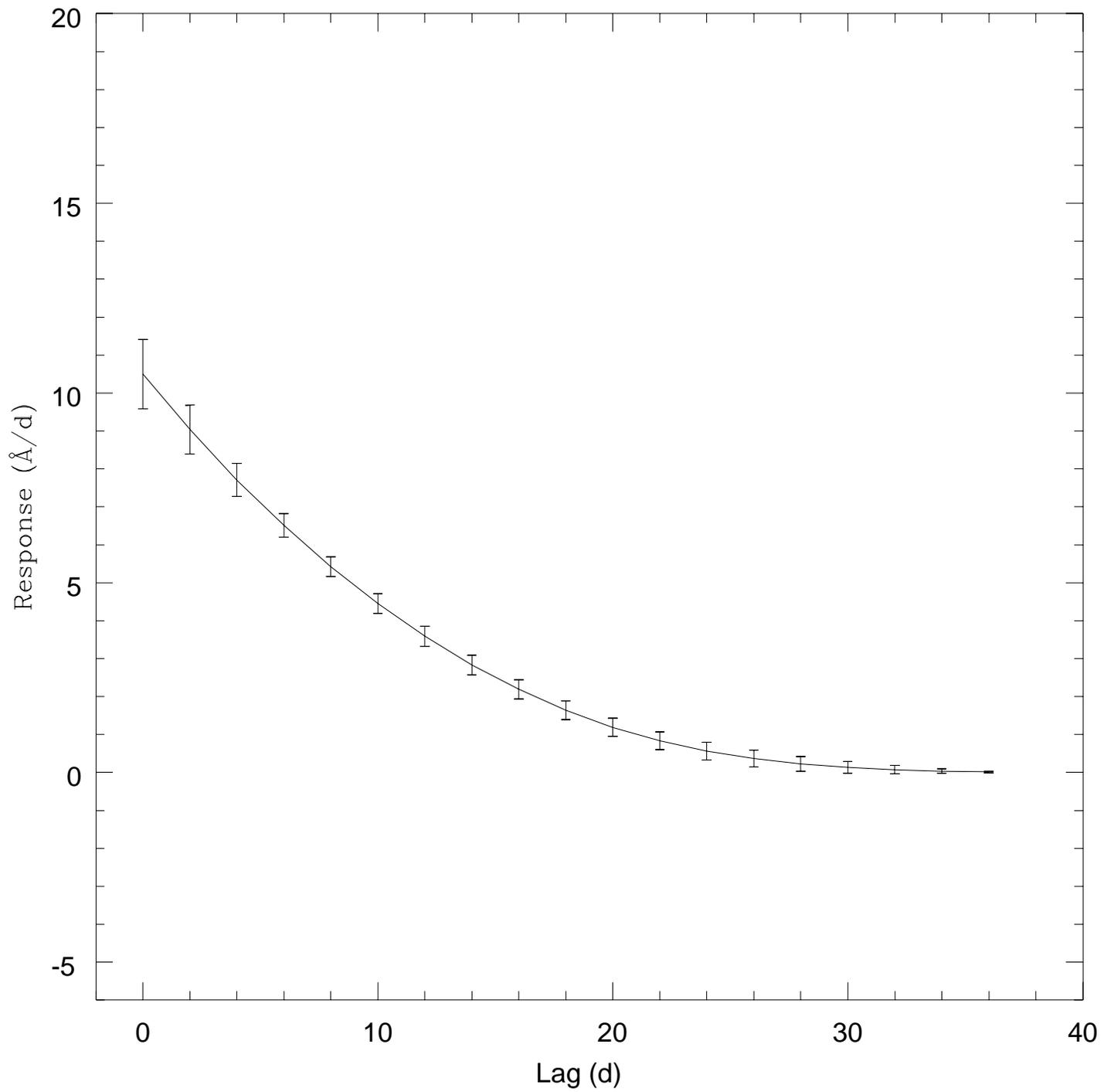
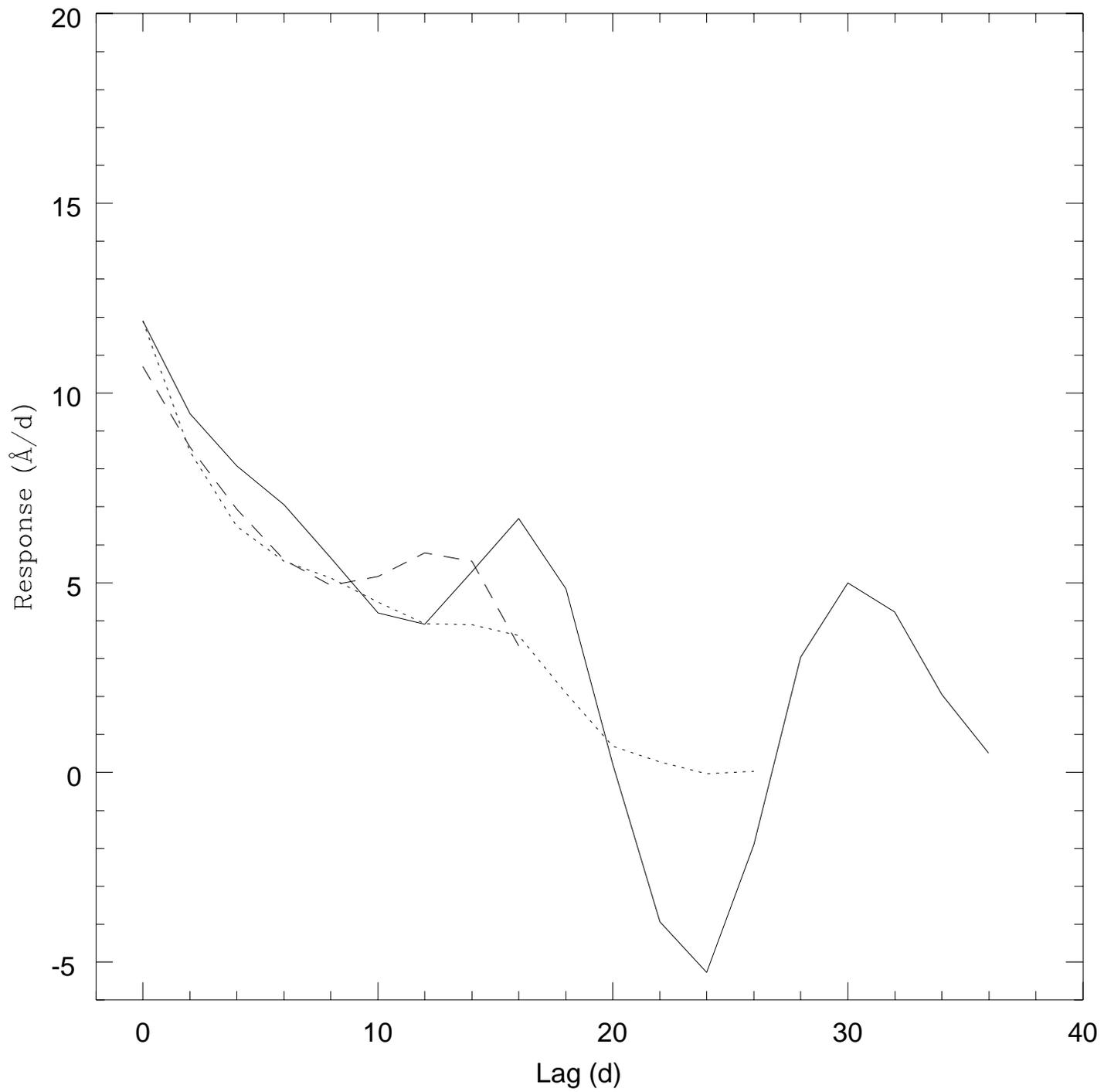
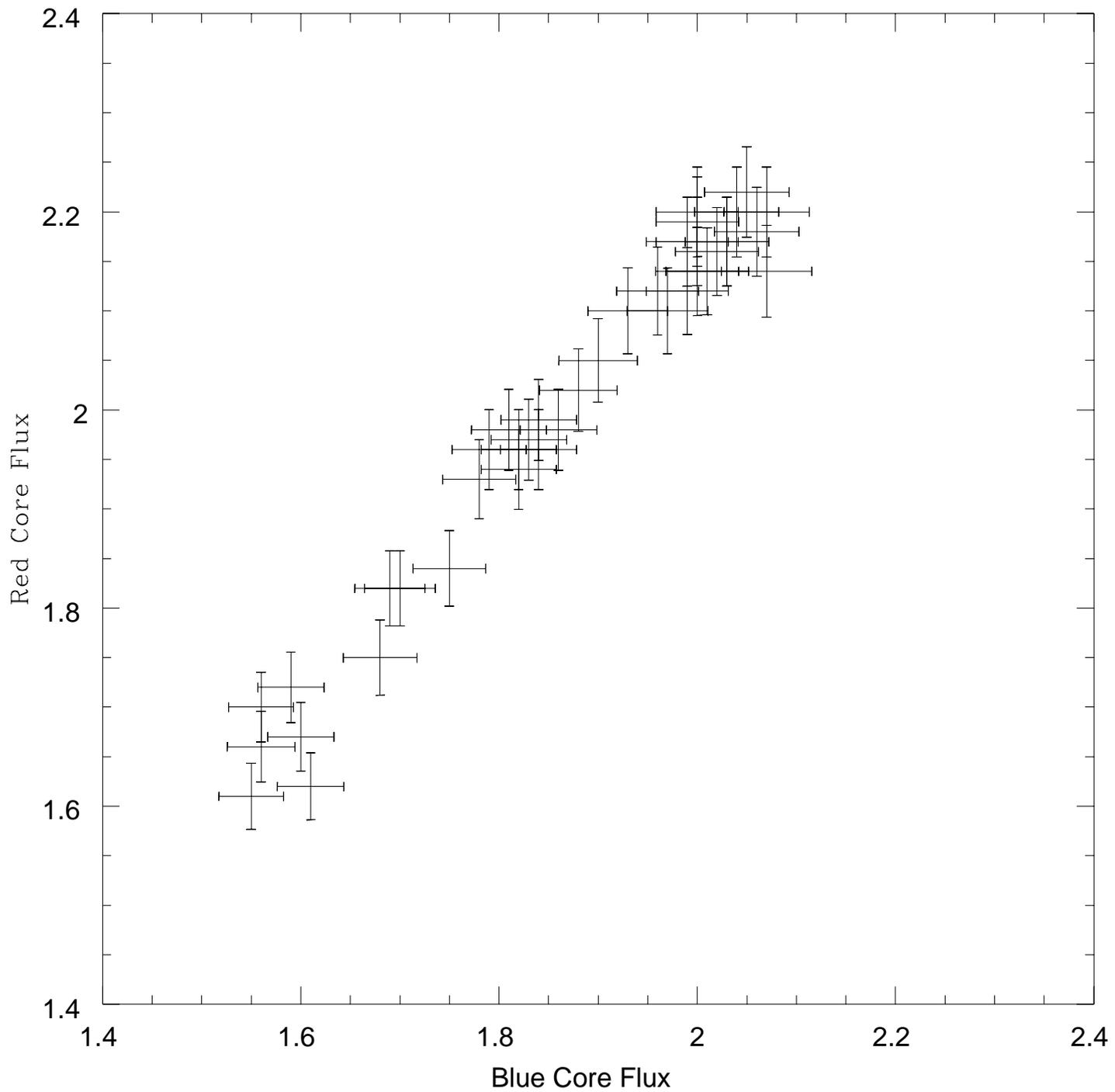
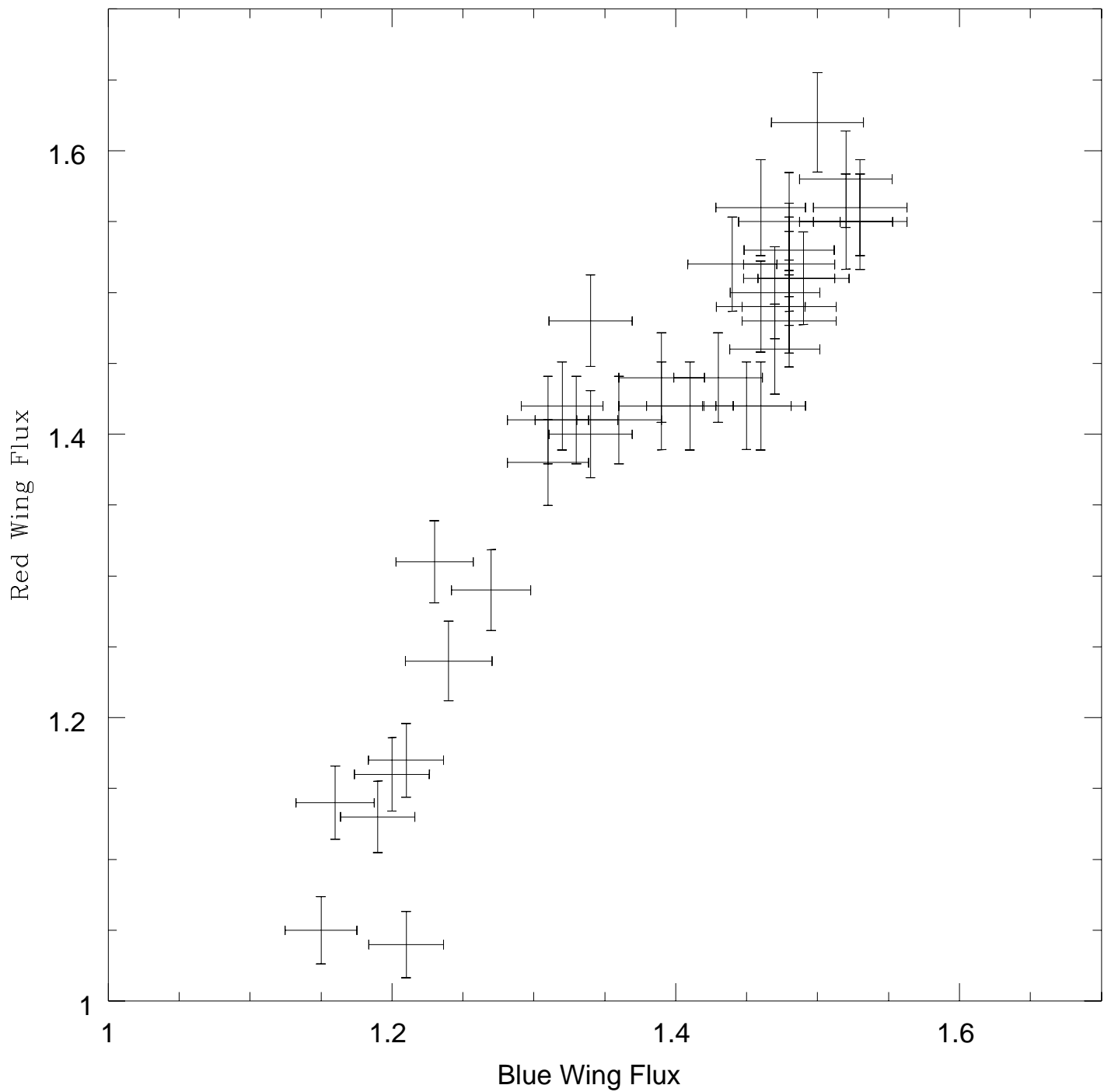
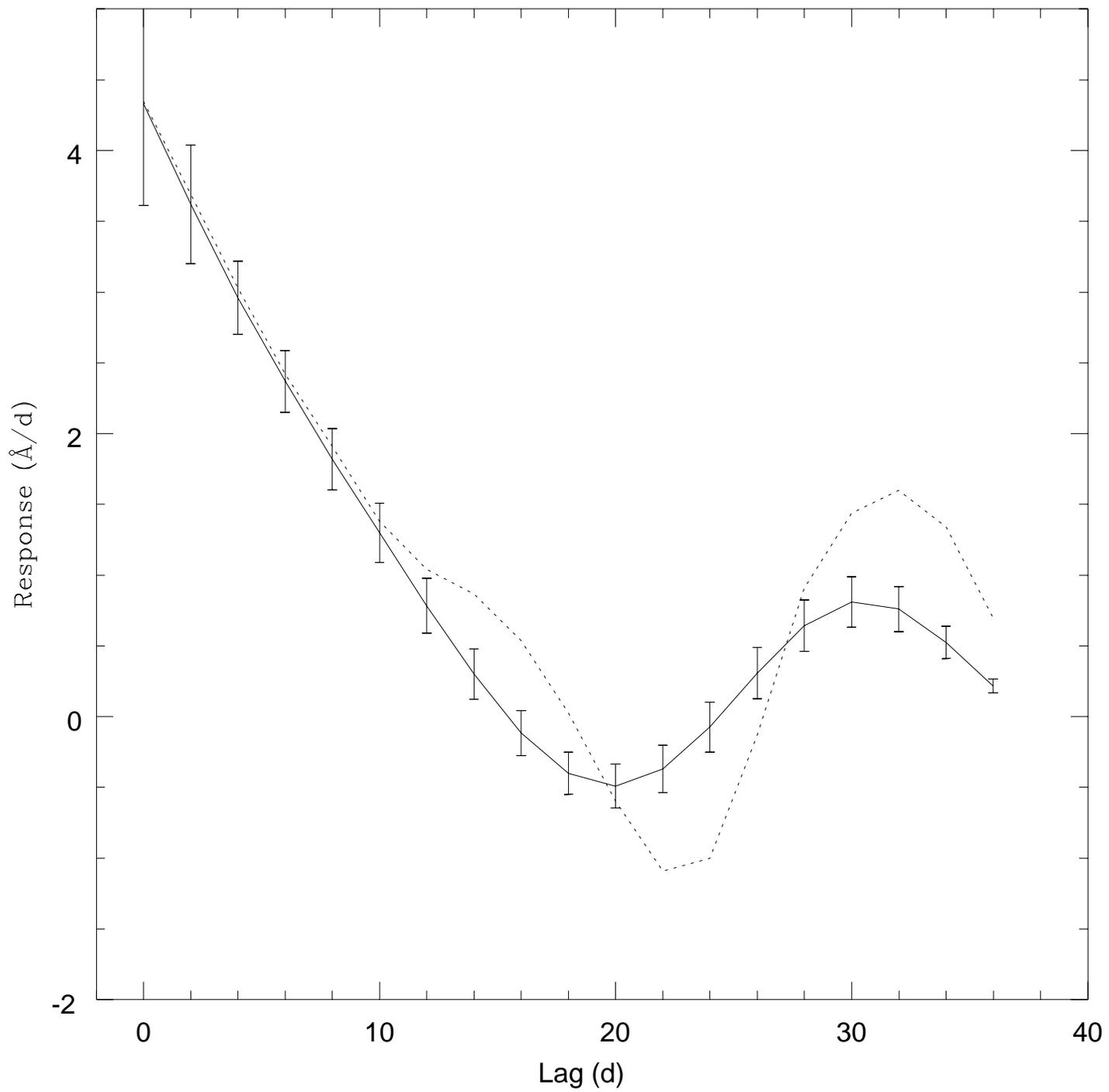
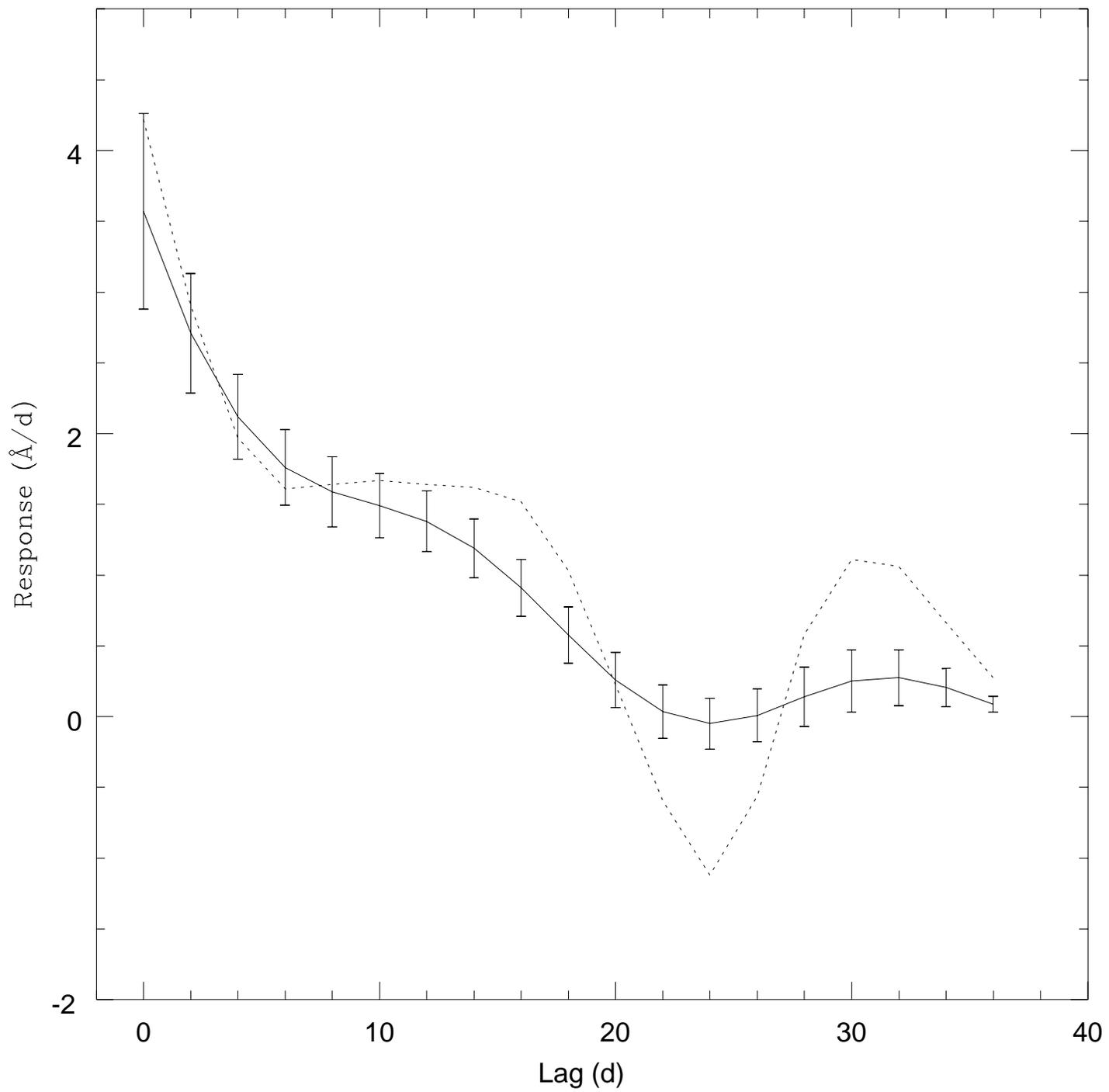
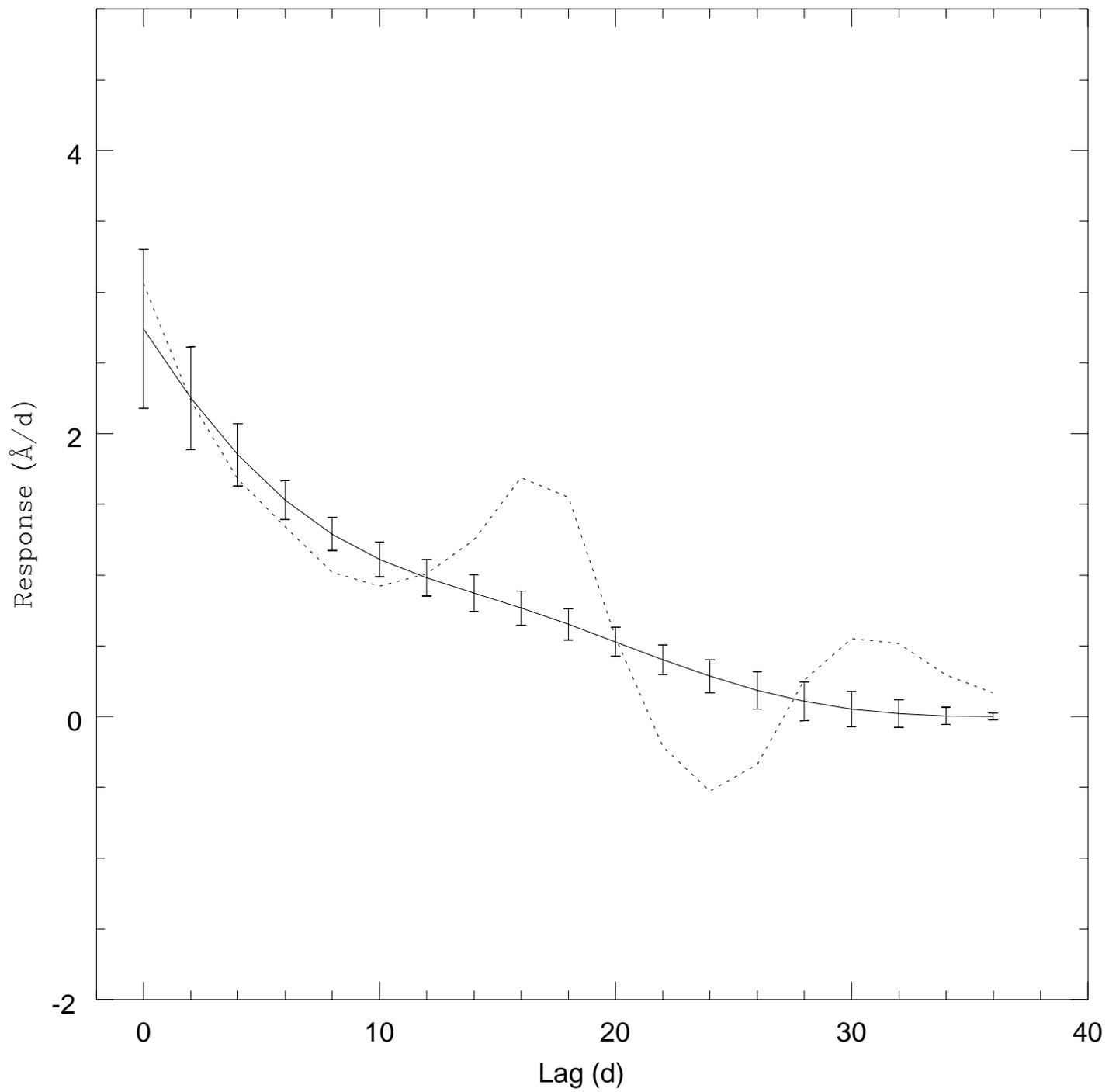
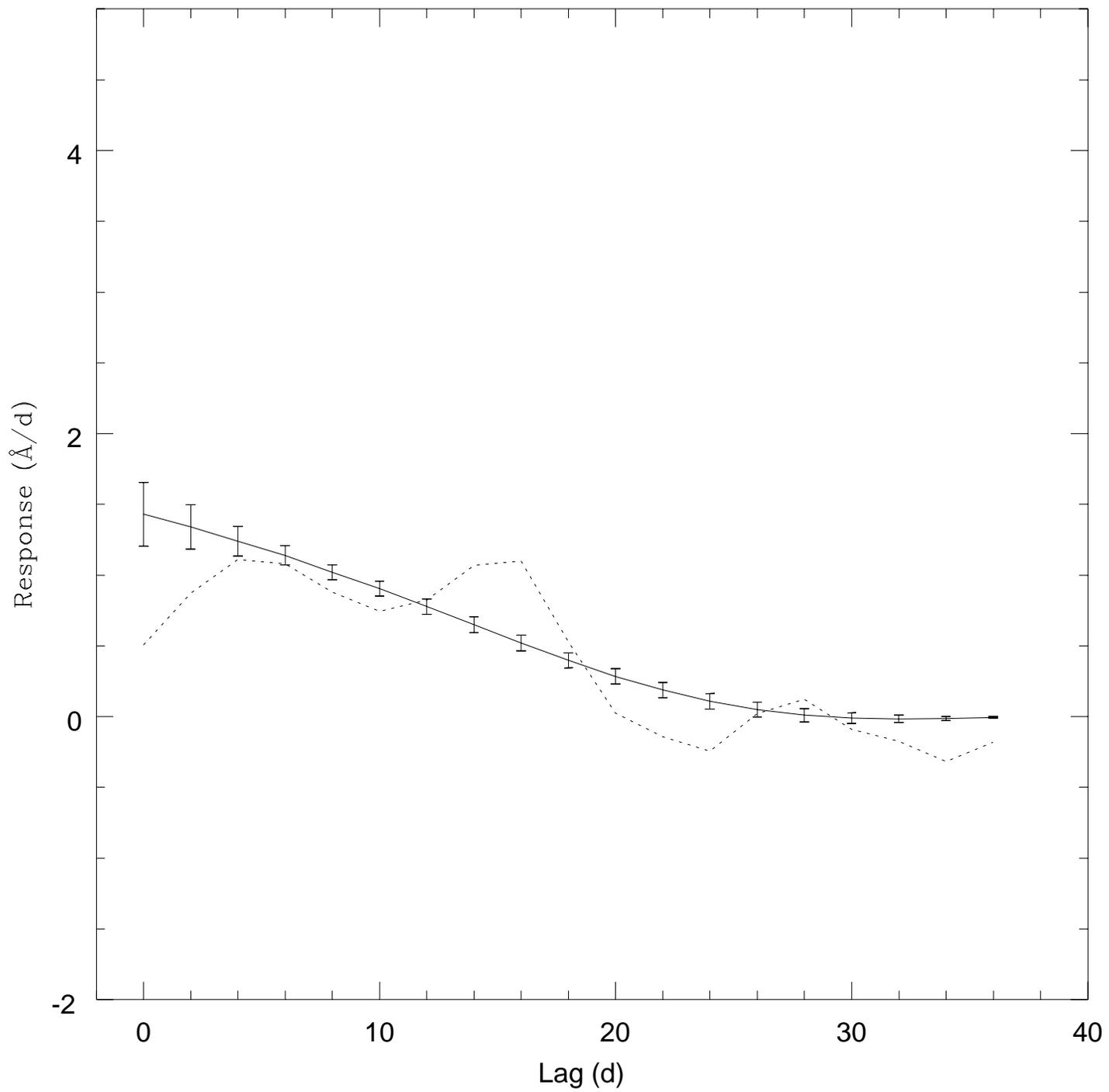
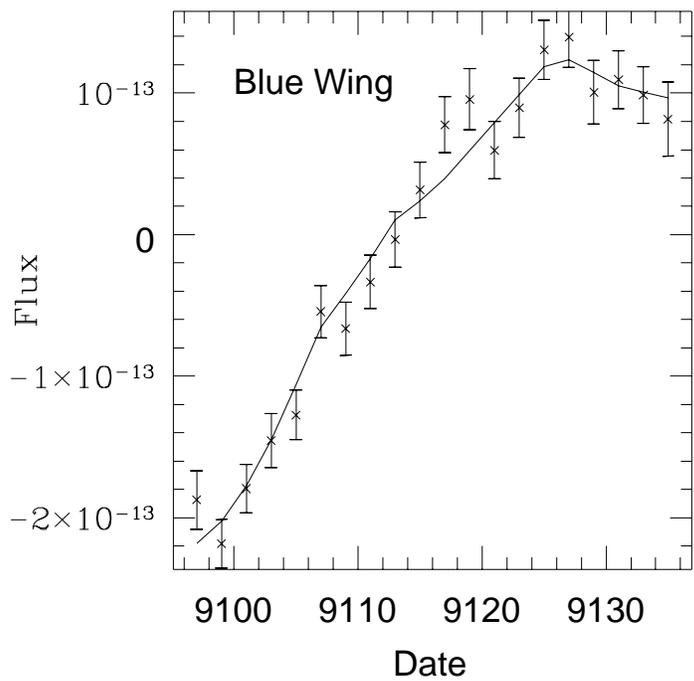
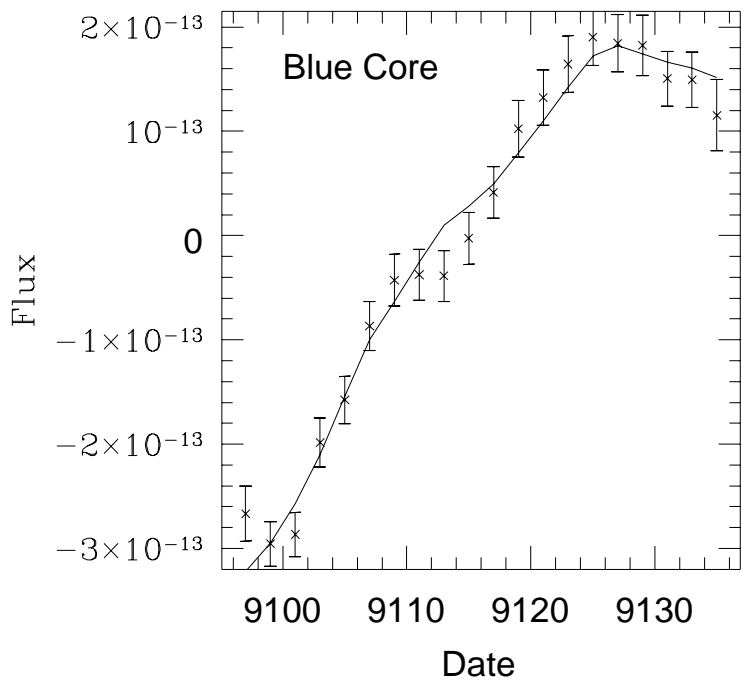
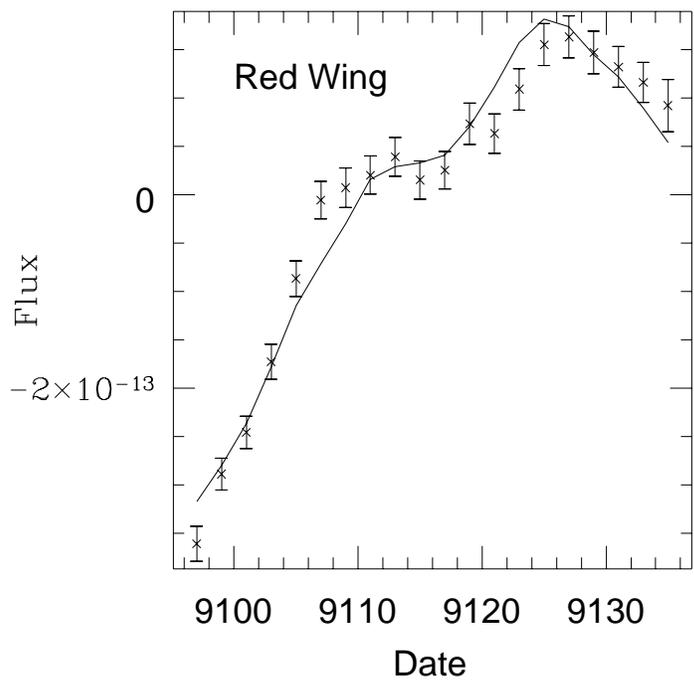
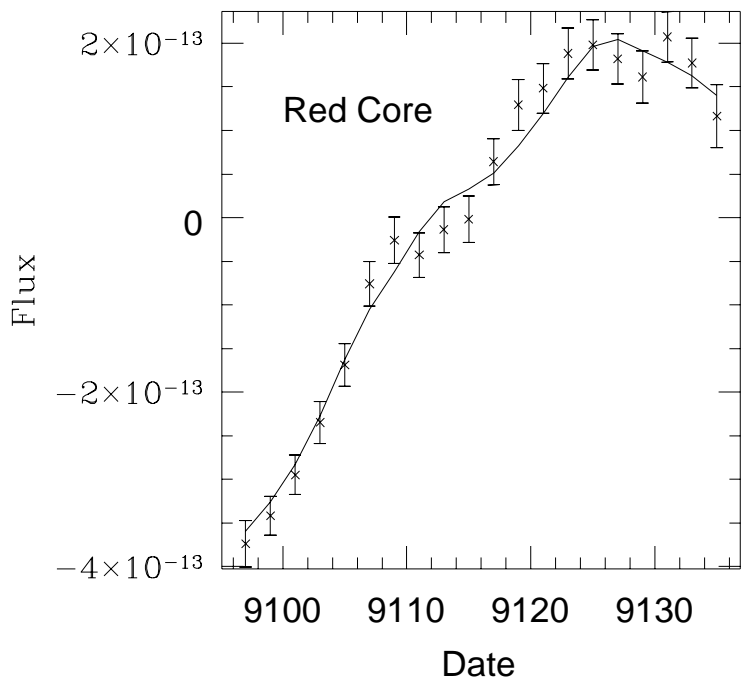
\end{document}